\newif\ifNotes\Notesfalse
\newif\ifNCH\NCHfalse

\documentclass[10pt, conference, compsocconf]{IEEEtran}

\usepackage[normalem]{ulem}
\usepackage{multirow}    
\usepackage{xspace}      
\usepackage{graphicx}
\usepackage{bbding}
\usepackage{xfrac} 
\usepackage{amssymb}
\usepackage[linesnumbered,ruled]{algorithm2e}
\makeatletter
\renewcommand{\@algocf@capt@plain}{above}
\makeatother 
\usepackage{algorithmicx}  
\usepackage{algpseudocode}
\usepackage[bookmarks=false,breaklinks=true,letterpaper=true,colorlinks,linkcolor=black,citecolor=blue,urlcolor=black]{hyperref}
\usepackage{booktabs}

\usepackage{tikz}

\newcommand{\name}{SoftTRR\xspace}
\newcommand{\mypara}[1]{\vspace{2pt}\noindent\textbf{{#1. }}}

\newcommand{\eat}[1]{}

\newcommand{\new}[1]{{\color{black}#1}\xspace}
\newcommand{\newpar}[1]{{\color{black}#1}\xspace}

\begin{document}

\title{\name: Protect Page Tables against RowHammer Attacks \\ using Software-only Target Row Refresh}

\author{\IEEEauthorblockN{Zhi Zhang\IEEEauthorrefmark{1}\IEEEauthorrefmark{2},
Yueqiang Cheng\IEEEauthorrefmark{1}\IEEEauthorrefmark{3},
Minghua Wang\IEEEauthorrefmark{4}, 
Wei He\IEEEauthorrefmark{5},
Wenhao Wang\IEEEauthorrefmark{6}, \\
Nepal Surya\IEEEauthorrefmark{2},
Yansong Gao\IEEEauthorrefmark{2}\IEEEauthorrefmark{7},
Kang Li\IEEEauthorrefmark{4},
Zhe Wang\IEEEauthorrefmark{8},
Chenggang Wu\IEEEauthorrefmark{8}
}
\IEEEauthorblockA{\IEEEauthorrefmark{1}
Both authors contributed equally to this work}
\IEEEauthorblockA{\IEEEauthorrefmark{2}
Data61, CSIRO, Australia Email: \{zhi.zhang,surya.nepal\}@data61.csiro.au}
\IEEEauthorblockA{\IEEEauthorrefmark{3}NIO Security Research Email: yueqiang.cheng@nio.io}
\IEEEauthorblockA{\IEEEauthorrefmark{4}Baidu Security Email: \{wangminghua01,kangli01\}@baidu.com}
\IEEEauthorblockA{\IEEEauthorrefmark{5}SKLOIS, Institute of Information Engineering, CAS and \\ 
School of Cyber Security, University of Chinese Academy of Sciences Email: hewei@iie.ac.cn}
\IEEEauthorblockA{\IEEEauthorrefmark{6}Institute of Information Engineering, CAS Email: wangwenhao@iie.ac.cn}
\IEEEauthorblockA{\IEEEauthorrefmark{7}NanJing University of Science and Technology, China Email: yansong.gao@njust.edu.cn}
\IEEEauthorblockA{\IEEEauthorrefmark{8}Institute of Computing Technology, Chinese Academy of Sciences Email: \{wangzhe12,wucg\}@ict.ac.cn}}

\maketitle

\begin{abstract}

Rowhammer attacks that corrupt level-1 page tables to gain kernel privilege are the most detrimental to system security and hard to mitigate. 
However, recently proposed software-only mitigations are not effective against such kernel privilege escalation attacks.

In this paper, we propose an effective and {practical} software-only defense, called \name, to protect page tables from all existing rowhammer attacks on x86. The {key idea of} \name is to refresh the rows occupied by page tables when a suspicious rowhammer activity is detected.
\name is motivated by DRAM-chip-based target row refresh (ChipTRR) but eliminates its main security limitation (i.e., ChipTRR tracks a limited number of rows and thus can be bypassed by many-sided hammer~\cite{frigo_trrespass_2020}).
Specifically, \name protects an unlimited number of page tables by tracking memory accesses to the rows that are in close proximity to page-table rows and refreshing the page-table rows {once} the tracked access count exceeds a pre-defined threshold.   
We implement a prototype of \name as a loadable kernel module, and evaluate its security effectiveness, performance overhead, and memory consumption. 
The experimental results show that \name protects page tables from real-world rowhammer attacks {and incurs} small performance overhead as well as memory cost.

\end{abstract}


\IEEEpeerreviewmaketitle

\section{Introduction}\label{sec:intro}
Rowhammer is a software-induced dynamic random-access memory (DRAM) vulnerability that frequently accessing (i.e., hammering) DRAM aggressor rows can induce bit flips in neighboring victim rows.
An attacker can hammer aggressor rows to corrupt different types of sensitive objects on victim rows without access to them, breaking memory management unit (MMU)-based memory protection,
achieving privilege escalation~\cite{seaborn2015exploiting, cheng2018still,zhang2019telehammer} or leaking sensitive information~\cite{bosman2016dedup,kwong2020rambleed}.
Of the many sensitive objects that have been corrupted by the rowhammer attacks, page table corruption is the most detrimental to system security, making kernel privilege escalation attacks the mainstream~\cite{wu2018CAT}. 
{To date,} kernel privilege escalation attacks~\cite{seaborn2015exploiting,gruss2016rowhammer, van2016drammer,xiao2016one, cheng2018still, zhang2019telehammer} focus on corrupting level-1 page table entry (L1PTE) and some of them have been demonstrated to gain kernel privilege from unprivileged applications~\cite{seaborn2015exploiting, cheng2018still,zhang2019telehammer}, or even from JavaScript in webpages~\cite{gruss2016rowhammer}.

Multiple software-only mitigation schemes~\cite{brasser17can,wu2018CAT,konoth2018zebram} can be used to mitigate the kernel privilege escalation attacks. Compared to hardware defenses~\cite{DDR4, lpDDR4,son2017making, lee2019twice}, software-only schemes have the appeal of compatibility with existing hardware, allowing better deployability. However, existing software-only mitigations require modifications to memory allocator and they
are {not effective against all the kernel privilege escalation attacks}. 
Specifically, CATT~\cite{brasser17can} and CTA~\cite{wu2018CAT} are vulnerable to a recent privilege escalation attack (PThammer~\cite{zhang2019telehammer}) that targets L1PTE.
ZebRAM~\cite{konoth2018zebram} assumes that bit flips occur in a victim row that is one-row from hammered aggressor row(s), making itself unable to defend against (kernel privilege escalation) rowhammer attacks where a victim row is no less than 2-row from the hammered rows~\cite{kim2020revisiting,zhang2019telehammer}. 
To this end, we ask:

\begin{center}
  \emph{Is there an effective and practical software-only defense that protects page tables against rowhammer attacks?}
\end{center}

\mypara{Our Contributions} 
{In this paper, we provide a positive answer to the question.} 
{We propose} a new software-only defense that defends against all \new{existing} kernel privilege escalation attacks on x86, called \name. 
\name is motivated by a hardware defense (ChipTRR~\cite{DDR4,lpDDR4}). 
ChipTRR is designed to count rows' activations and refreshing adjacent rows to suppress bit flips if the activation counts reach a pre-defined threshold. ChipTRR was believed to eliminate the rowhammer effect in present-day DDR4-based systems, until it was completely circumvented by~\cite{frigo_trrespass_2020}. 

We observe that the root cause of failure of ChipTRR is that it tracks a limited number of rows. Thus, bit flips are still possible when multiple rows are being hammered and the number of hammered rows is larger than the tracked rows (i.e., \emph{many-sided hammer}~\cite{frigo_trrespass_2020}). \name addresses this limitation by monitoring and tracking all rows neighboring (victim) rows containing page tables. 
\name leverages MMU-enforced virtual memory subsystem to frequently track memory accesses to any rows adjacent to page-table rows, and refreshes page-table rows when necessary, making \name effective in preventing rowhammer from breaking page table integrity.

Specifically, MMU is an essential component of modern processors that supports OS kernel to enforce memory isolation. 
With the assistance from MMU, the kernel, configures page tables, mediates every memory access from user space, and captures any unauthorized access that triggers a hardware exception.
On top of that, the kernel can capture the memory access where relevant page tables have an unused \texttt{rsrv} bit set. 
With this {observation}, \name uses the kernel as the root of trust and frequently configures page tables with the \texttt{rsrv} bit set to track memory accesses to rows that neighbor rows of page tables. When the tracked memory-access counters reach a pre-determined limit, corresponding page-table rows will be refreshed. 
{By} \name's design, an adjacent or neighboring row can be multiple-row from a page-table row, thus voiding the above assumption of one-row-distance between victim and aggressor rows made by ZebRAM~\cite{konoth2018zebram}. 
In our implementation, the adjacent rows are up to 6-row away from the aggressor rows, the largest row distance that has been observed so far~\cite{kim2020revisiting}.


Our prototype implementation of \name is a loadable kernel module (LKM) without any modification to the kernel. The LKM has about 2320 source lines of code and it has been deployed into three Linux systems where underlying hardware have either DDR3 or DDR4 modules.
We evaluated \name-deployed systems in terms of security effectiveness, performance, memory consumption and robustness.
The experimental results show that \name is effective in mitigating kernel privilege escalation attacks. Besides, \name incurs low overhead on the tested benchmarks and its memory consumption is within hundreds of KiB in a real-world use case of LAMP (i.e., Linux, Apache, Mysql and PHP). We also {validate} the robustness of a \name-enabled system using system-call stress tests, results of which show that the system runs as stable as a vanilla system. 

In summary, the main contributions are as follows:

\vspace{2pt}\noindent$\bullet$  We {introduce} \name to protect page tables against rowhammer attacks. 
Compared to prior works, \name is an effective and practical software-only mitigation scheme.

\vspace{2pt}\noindent$\bullet$  We implement a lightweight \name prototype to collect page tables, track memory access, and refresh target page tables by leveraging MMU and OS kernel features.

\vspace{2pt}\noindent$\bullet$  We evaluate \name's effectiveness against 3 representative rowhammer attacks, its performance overhead and memory consumption. The experimental results show that \name successfully protects page tables against the attacks, and {incurs} negligible overhead and memory cost. 


The rest of the paper is structured as follows.
In \autoref{sec:bkgd}, we introduce DRAM, rowhammer vulnerability and rowhammer defenses. 
In \autoref{sec:srr} and \autoref{sec:impl}, we {present} the design and implementation of \name. 
\autoref{sec:eva} and \autoref{sec:perf_eva} evaluate \name's security effectiveness and performance impacts, respectively.
We discuss and conclude this paper in \autoref{sec:dis} and \autoref{sec:conclusion}.


\section{Background and Related Work}\label{sec:bkgd}
In this section, we first describe 
DRAM and its address mapping. We then present the rowhammer vulnerability as well as its hardware and software defenses (\cite{mutlu2019rowhammer} provides a detailed survey of the rowhammer).

\eat{
\subsection{Address Translation} \label{sec:page-table-walk}
MMU enforces virtual memory primarily by the means of paging mechanism. Paging on the x86-64 platform usually uses four levels of page tables to translate a virtual address to a physical address. As such, virtual-address bits are usually divided into 4 parts as follows.

The bits 39--47 are used to index a page map level table entry (PML4 or level-4 page table). The physical base address of the PML4 is stored in the control register of CR3. 
The bits 30--38 are used to index a page directory pointer table entry (PDPT or level-3 page table). 
The physical base address of the PDPT comes from the PML4 entry.
The bits 21--29 are used to index a page directory table entry (PD or level-2 page table) and the physical base address of the PD comes from the PDPT entry. 
The bits 12--20 are used to index a page table entry (PT or level-1 page table) and the physical base address of PT comes from the PD entry.
Now the indexed PT entry points to a physical page and the rest bits, i.e., 0--11, are page offset. 
If the physical address is within a \emph{huge page}, either two or three levels of page tables are needed to translate a virtual address within a huge page of 1\,GiB or 1\,2MiB. 
In order to facilitate the address translation, Translation Look-aside Buffer (TLB) is introduced to cache the address translations while cache is involved to store the accessed data as well as the page table entries of all levels.
}

\subsection{DRAM}
The main memory of most modern computers uses DRAM. Memory modules are usually produced in the form of dual inline memory module (DIMM), where both sides of the memory module have separate electrical contacts for memory chips. Each memory module is directly connected to the CPU's memory controller through one of the two channels. Logically, each memory module consists of two ranks, corresponding to its two sides, and each rank consists of multiple banks. A bank is structured as arrays of memory cells with rows and columns. 

Every cell of a bank stores one bit of data whose value depends on whether the cell is electrically charged or not. A row is a basic unit for memory access. Each access to a bank ``opens'' a row by transferring the data from all of the cells of the row to the bank's \emph{row buffer} that acts as a cache for the most recently accessed row. This operation discharges all the cells of the row. To prevent data loss, the row buffer is copied back into the cells, thus recharging the cells. Consecutive access to the same row is fulfilled from the row buffer,  while accessing another row flushes the row buffer.
As the charge stored in the cell disperses over time, every cell's charge must be restored or refreshed periodically in a specified time period. The typical refresh period is 64\,milliseconds (ms).

\mypara{DRAM Address Mapping}
The memory controller decides how physical-address bits are mapped to a DRAM address. 
A DRAM address refers to a 3-tuple of \emph{bank, {row}, {column}} (DIMM, channel, and rank are included into the \emph{bank} tuple field). As this mapping is not publicly documented on the Intel processor platform, it has been reverse-engineered by multiple works~\cite{seaborndram,pessl2016drama,xiao2016one,wang2020dramdig}. Most of them exploited a timing side channel~~\cite{moscibroda2007memory} to uncover the mapping, that is, accessing two virtual addresses that reside in different rows of the same bank leads to higher access latency when compared to accessing the addresses that are in different banks or in the same row of the same bank. 


\subsection{Rowhammer Vulnerability}
Kim et al.~\cite{kim2014flipping} are the first to perform a large scale study of rowhammer on DDR3 modules, results of which have shown that the vulnerability can be triggered by software accesses, that is, frequently accessing rows of $i+1$ and $i-1$  (i.e., aggressor rows) cause bit flips (i.e., charge leakage) in row $i$ (i.e., victim row). 

There are four hammer patterns. 
First, \emph{double-sided hammer} refers to a case where two adjacent rows of the victim row are hammered simultaneously, which is the most effective hammer pattern in inducing bit flips on DDR3 modules~\cite{seaborn2015exploiting}. 
Second, \emph{single-sided hammer} randomly picks two aggressor rows in the same bank and hammers them~\cite{seaborn2015exploiting}. 
Third, \emph{one-location hammer} selects a single aggressor row for hammer. This hammer pattern only applies to certain systems where the DRAM controller employs an advanced policy to optimize performance~\cite{gruss2017another}.
Last, \emph{many-sided hammer} chooses more than two aggressor rows within the same bank for hammer. The aggressor rows are usually separated by one row and two out of them are adjacent to the victim row~\cite{frigo_trrespass_2020}.

\subsection{Rowhammer Defenses}
\mypara{Hardware Solutions}
Existing hardware solutions employed by the industry can be summarized into three categories. The first is to decrease the DRAM refresh period~\cite{kim2014flipping} to refresh all DRAM rows more frequently. For instance, three computer manufacturers (i.e., HP~\cite{HP}, Lenovo~\cite{LENOVO} and Apple~\cite{Apple}) deployed firmware updates to decrease the refresh period from 64\,ms to 32\,ms. 
However, \emph{clflush-free} rowhammer attacks~\cite{aweke2016anvil} can still induce bit flips in the refresh period of 32\,ms. Decreasing the refresh period by more than 7x can make the rowhammer impossible but it will impose unacceptable overhead to the systems~\cite{kim2014flipping}.
The second one is proposed by Intel~\cite{intelecc} that leverages Error Correcting Code (ECC) memory to correct single-bit errors and detect double-bit errors. However, ECC has been reverse engineered and is vulnerable to rowhammer~\cite{cojocar2019exploiting}. 
The last is to track row's activation count and various approaches have been proposed~\cite{kim2014flipping, seyedzadeh2016counter, son2017making, seyedzadeh2018mitigating,rowhammer,DDR4,lpDDR4,lee2019twice,park2020graphene,yauglikcci2021blockhammer}. Among them, ChipTRR~\cite{lpDDR4,DDR4} was adopted by recent DDR4 manufacturers but it has been defeated by TRRespass~\cite{frigo_trrespass_2020}. \new{None of other approaches are widely deployed due to their limitations (e.g., significant area cost or performance downsides)~\cite{bennettpanopticon}.}

\mypara{Software Defenses}
Software defenses include both mitigation and detection techniques. As sensitive data is required to be within victim rows for exploitation, existing mitigation techniques modify memory allocator and enforce DRAM-aware memory isolation at different granularity~\cite{brasser17can, van2018guardion, tatar2018throwhammer, wu2018CAT, bock2019rip,konoth2018zebram}.  
CATT~\cite{brasser17can} implements DRAM isolation between user and kernel memory. 
CTA~\cite{wu2018CAT} provides a dedicated DRAM region for level-1 page tables. ZebRAM~\cite{konoth2018zebram} isolates rows of sensitive data in a zebra pattern.
These defenses can prevent page tables from being hammered. 
Albeit on different hardware, SoftTRR has an averaged overhead of 0.8\% on \texttt{SPECint 2006}, similar to that of CATT~\cite{brasser17can} and CTA~\cite{wu2018CAT}. However, ZebRAM has a much higher overhead of 4\%--5\%.
ALIS~\cite{tatar2018throwhammer} isolates DMA memory to prevent the remote rowhammer attack (i.e., Throwhammer~\cite{tatar2018throwhammer}) targeting a key-value user application.
RIP-RH~\cite{bock2019rip} provides DRAM isolation for local user processes.

Anvil~\cite{aweke2016anvil} utilizes CPU performance counters to monitor cache miss rate and detects a rowhammer attack, as typical rowhammer attacks incur frequent cache misses. However, Anvil is prone to false positives and/or false negatives~\cite{wu2018CAT,brasser17can}. Besides,
its current implementation cannot detect the PThammer attack~\cite{zhang2019telehammer}.
The other detection technique is RADAR~\cite{zhangleveraging}. As 
rowhammer attacks exhibit recognizable rowhammer-correlated sideband patterns in the spectrum of the DRAM clock signal, RADAR leverages peripheral customized devices to capture and analyze the electromagnetic signals emitted by a DRAM-based system. 


\begin{figure*}[hbt!]
\centering
\includegraphics[width=\textwidth]{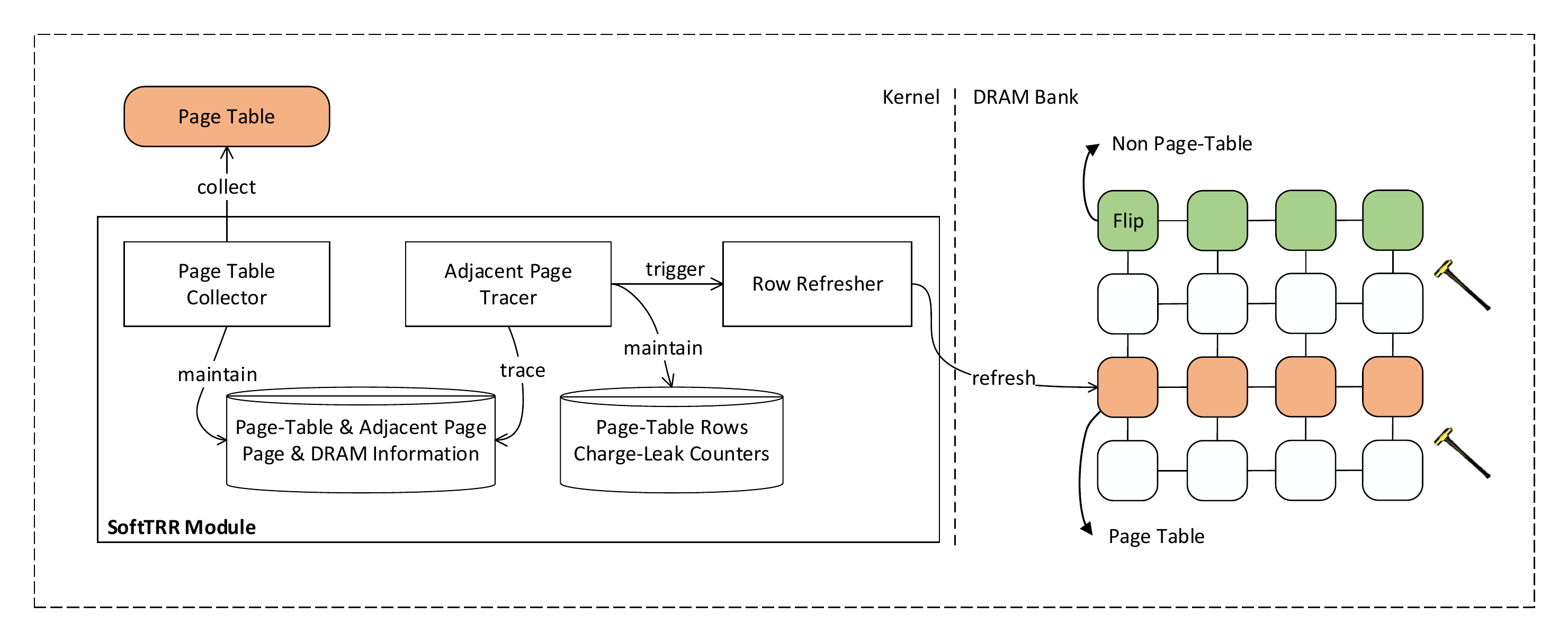}
\caption{\name Overview. \name is a kernel module and has three main components. \emph{Page Table collector} maintains information about page-table pages and their adjacent pages in close proximity. \emph{Adjacent Page Tracer} traces access to the maintained adjacent pages and updates charge-leak counters for relevant rows of page-table pages. When the counters reach a pre-determined limit, \emph{Row Refresher} is triggered to refresh desired rows hosting page-table pages. In comparison, \new{non-page-table} rows (highlighted in green) are vulnerable to bit flips.} 
\label{fig:soft_srr}
\end{figure*}

\section{\name: Software-only Target \\ Row Refresh}
\label{sec:srr}
We discuss threat model and assumptions in \autoref{sec:threatmodel}, design principles in \autoref{sec:principle} and design overview in \autoref{sec:overview}. \autoref{sec:impl} describes implementation details. 

\subsection{Threat Model and Assumptions}
\label{sec:threatmodel}

Our primary goal is to protect page tables and guarantee that an adversary cannot corrupt them to gain kernel privilege through rowhammer on x86 architectures.
In our implementation of \name, we focus on protecting level-1 page tables (L1PTs), the same goal as in CTA~\cite{wu2018CAT}, because all existing page-table-oriented rowhammer attacks aim at corrupting L1PTs.
Even when higher levels of PTs are corrupted, they are hard to be exploited (see details in CTA~\cite{wu2018CAT}).
In spite of that, \name can be extended to protect other levels of page tables and we discuss it in \autoref{sec:dis}.


We assume the kernel as the root of trust, and the kernel module implementing \name is well protected. 
We consider threats coming from both local adversaries and remote adversaries. A local adversary resides in a low privilege user process and thus can execute arbitrary code within her privilege boundary. 
A remote adversary stays outside by launching an attack, e.g., through a website with JavaScript. 

The DRAM address mappings and in-DRAM address remappings can be reverse-engineered using prior works~\cite{xiao2016one,pessl2016drama,wang2020dramdig,cojocar2020we} and they are assumed to be available. 
Besides, previous software-only rowhammer defenses~\cite{brasser17can,wu2018CAT,bock2019rip, konoth2018zebram} consider that hammering $row_i$ only affects $row_{i+1}$ and $row_{i-1}$, which however is not consistent with a recent work by Kim et al.~\cite{kim2020revisiting}. Particularly, they performed a comprehensive study of 1580 DRAM chips (300 DRAM modules in total) from three major DRAM manufacturers and found that bit flips can occur in rows that are up to 6-row away from the hammered $row_i$. 
\name by design protects rows of page tables from being flipped by rows that are \texttt{N}-row away and its current implementation allows that the distance between an adjacent row and an L1PT row ranges from 1-row to 6-row, the largest row distance observed by Kim et al~\cite{kim2020revisiting}.




\subsection{Design Principles}
\label{sec:principle}
\name follows the security and practicality design principles described below.
The security principle is to guarantee \name can defend against all existing rowhammer attacks targeting page tables. The practicality principles aim to make \name applicable to real-world systems.

\vspace{2pt}\noindent$\bullet$  \textbf{DP1:} \name should be effective in protecting ALL page tables. 
Without this completeness guarantee, an attacker can gain kernel privilege by compromising the integrity of page tables that are not protected by \name.

\vspace{2pt}\noindent$\bullet$  \textbf{DP2:} \name should be compatible with OS kernels. It neither modifies/adds kernel source code nor breaks kernel code integrity through binary instrumentation, which hinders its adoption in practice. 

\vspace{2pt}\noindent$\bullet$  \textbf{DP3:} \name should \newpar{have} small performance overhead to a protected system. 

\subsection{Design Overview}\label{sec:overview}
\name, residing in the kernel space, collects all page tables, and monitors their entire life cycle from page-table creation to page-table release.
For each collected page-table page, \name identifies all its adjacent pages in DRAM and traces memory accesses to the adjacent pages. 
Thus, \name is aware of which adjacent page is accessed.
When the traced access count reaches a pre-determined limit, \name knows which page-table page is at the risk of being flipped and promptly refreshes the page (satisfying \textbf{DP1}). 

All \new{existing} software-only mitigation techniques (see \autoref{sec:bkgd}) deeply hack into the memory allocator to become DRAM-aware and add extra allocation/deallocation constraints. Unlike them, \name only acquires offline domain knowledge (e.g., DRAM address (re)mappings of physical addresses), without requiring a new memory allocator or changing legacy allocator logic (satisfying \textbf{DP2}).

\name configures page tables to trace memory access{es} to those adjacent pages. Thus, the access to an adjacent page raises a hardware exception, which is captured by \name for the tracing purpose. If no such access occurs, no overhead is introduced. 
Thus, the accesses to non-adjacent pages are at full speed, isolating the performance overhead caused by the accesses to adjacent pages (satisfying \textbf{DP3}).

As shown in \autoref{fig:soft_srr}, \name has three critical components. 
\emph{Page Table Collector} actively collects all page tables and maintains their page and DRAM information. It also collects and maintains \emph{adjacent pages}.
Besides being accessible to unprivileged users, a page is considered as adjacent when itself or its corresponding page-table page is adjacent to (\texttt{N}-row from) another page-table page. This is based on an observation from Zhang et al.~\cite{zhang2019telehammer}. In particular, rowhammer attacks corrupting page tables are classified into two categories. For \emph{explicit} attacks \cite{seaborn2015exploiting,cheng2018still}, they require attacker-accessible memory adjacent to L1PT pages. For \emph{implicit} attacks~\cite{zhang2019telehammer}, they only need mutual adjacency among L1PT pages.

\emph{Adjacent Page Tracer} keeps a close watch over memory accesses to collected adjacent pages, and maintains a charge-leak counter for a row where a page-table page reside{s}. 
If any one row of adjacent pages has been accessed, the charge-leak counters of nearby page-table rows are updated accordingly, indicating that the page-table rows leak charge once. 

\emph{Row Refresher} remains dormant if charge-leak counters do not reach a pre-determined limit. {If yes}, a rowhammer attempt is believed to be taking place and the above tracer triggers row refresher, which will promptly refreshes desired rows whose charge-leak counters reach the limit.


\begin{table*}[htp!]
\centering
\begin{tabular}{llll}
\toprule
\multicolumn{1}{c}{\multirow{1}{*}{\new{\textbf{Data Structures}}}} & 
\multicolumn{2}{c}{\textbf{Main Fields in A Node}}  & \multicolumn{1}{c}{\textbf{Descriptions}}  \\\hline

\multicolumn{1}{c}{\texttt{pt\_rbtree}} & \multicolumn{2}{c}{PPN \new{(key)}} & {A unique page frame number of an L1PT page.} \\\hline

\multicolumn{1}{c}{\texttt{adj\_rbtree}} & \multicolumn{2}{c}{PPN \new{(key)}}  & {A unique page frame number of an adjacent page.} \\\hline

\multicolumn{1}{c}{\multirow{4}{*}{\texttt{pt\_row\_rbtree}}} & \multicolumn{2}{c}{row\_index \new{(key)}}  & {A row index of one or more L1PT pages.}\\
 & \multicolumn{1}{c|}{\multirow{3}{*}{bank\_struct}} & {bank\_index} & {A bank index of one or more L1PT pages.} \\
 & \multicolumn{1}{c|}{} & {pt\_count} & {The number of L1PT pages that have the same indexes of bank and row.} \\
 & \multicolumn{1}{c|}{} & {leak\_count} & {The number of accesses to rows adjacent to a row of row\_index and bank\_index.} \\\hline

\multicolumn{1}{c}{\multirow{3}{*}{\texttt{pte\_ringbuf}}} & \multicolumn{2}{c}{pte}  & {A pointer to a page table entry relevant to an adjacent page.}\\
& \multicolumn{2}{c}{vaddr} & {A virtual address relevant to an adjacent page.}\\
& \multicolumn{2}{c}{mm}  & {A pointer to \texttt{mm\_struct} relevant to a process where \texttt{vaddr} resides.} \\
\bottomrule
\end{tabular}
\caption{Data structures used by \name.}
\label{tab:structres}
\end{table*}

\section{Implementation}
\label{sec:impl}
As stated in \autoref{sec:threatmodel}, \name implements L1PT protection and a row of adjacent pages can be up to 6-row away from a row of L1PT pages.
Our prototype implementation is a loadable kernel module (LKM) without modifications to the kernel. The LKM consists of around 2320 source lines of code and works with Ubuntu installation running a default Linux kernel \new{4.4.211}. 
Before we talk about the three aforementioned components of \name,
we first introduce important data structures as below. 

\subsection{Data Structures}
We reuse the kernel's red-black tree structure~\cite{redblack}, an efficient self-balancing binary search tree that guarantees searching in $\Theta({\log n})$ time ($n$ is the number of tree nodes).
As shown in \autoref{tab:structres},
we have three red-black trees and a ring buffer, i.e., \texttt{pt\_rbtree},  \texttt{adj\_rbtree}, \texttt{pt\_row\_rbtree} and \texttt{pte\_ringbuf}, respectively. 

Specifically, \texttt{pt\_rbtree} stores L1PT page information while \texttt{adj\_rbtree} stores information of pages that are adjacent to L1PT pages. For the two trees, a physical page number (PPN) is used as the node key and thus a new node will be allocated when information of a new L1PT page or adjacent page needs to be stored. 
Besides, \texttt{pt\_row\_rbtree} stores DRAM information about L1PT pages.
For this tree node, \texttt{row\_index} works as the node key and a node can have one or more bank structures (i.e., \texttt{bank\_struct}). One bank structure stores \texttt{bank\_index} that one or more L1PT pages own (e.g., multiple L1PT pages share the same row of the same bank). Also note that a page can span across multiple banks~\cite{wang2020dramdig} and thus an L1PT page can have multiple \texttt{bank\_struct}. 
\texttt{pt\_count} records the number of L1PT PPNs that are in the same row of the same bank. \texttt{leak\_count}, short for the charge-leak counter in \autoref{sec:overview}, stores the number of accesses to rows that are adjacent to a row of \texttt{row\_index} in the same bank. 

For a given DRAM module, we leverage a publicly available tool, called DRAMA\footnote{https://github.com/IAIK/drama} to reverse-engineer its DRAM address mapping, and embed the mapping into the kernel before acquiring a physical page's DRAM information.
We allocate each node of each tree using the slab allocator~\cite{bonwick1994slab}, which is an efficient memory management mechanism intended for the kernel's small object allocation compared to the buddy allocator. 

\texttt{pte\_ringbuf} stores information of \new{leaf} page table entries (PTEs) that are collected by adjacent page tracer (see \autoref{sec:tracer}). 
These PTEs point to either adjacent pages themselves or \emph{huge pages} containing adjacent pages. 
If the adjacent page is a 4\,KiB page, the PTE is an L1PT entry. If the adjacent page is part of a {huge page} (i.e., 2\,MiB or 1\,GiB), the PTE is either an L2PT entry or an L3PT entry. 
Each node of \texttt{pte\_ringbuf} is a structure that has three main fields also shown in \autoref{tab:structres}.
Particularly, \texttt{pte} is a pointer to the \new{leaf} PTE.
\texttt{vaddr} is a virtual address referring to an adjacent page or its corresponding huge page. \texttt{mm} is a pointer to a kernel structure (i.e., \texttt{mm\_struct}) about a process's address space where \texttt{vaddr} belongs.
The adjacent page tracer combines \texttt{vaddr} and \texttt{mm} to flush the TLB entry that stores the adjacent page's virtual-to-physical address mapping.


\subsection{Page Table Collector}~\label{sec:collector}

For processes that are already in the main memory before our module is loaded, page table collector enquires the list of \texttt{task\_struct} to find every existing process. It then performs page-table walk for every virtual page in each valid virtual memory area (VMA) of each user process to collect information of L1PT pages and their adjacent pages.
Specifically, \texttt{pt\_rbtree} and \texttt{pt\_row\_rbtree} store distinct L1PT pages, and their DRAM bank and row indexes, respectively. 
To build \texttt{adj\_rbtree}, the collector finds out all user pages that are adjacent to L1PT pages in DRAM. It also selects all L1PT pages from \texttt{pt\_rbtree} that are adjacent to each other and puts all PPNs pointed by selected L1PT pages' valid entries into \texttt{adj\_rbtree}.
For free pages that are adjacent to L1PT pages and allocated for use later (e.g., a free page is allocated and mapped to the user space right after the collector finishes collecting all adjacent pages), the adjacent page tracer handles them appropriately (see \autoref{sec:tracer}).

For L1PT pages that are dynamically allocated or freed after the above collection, we perform dynamic inline hooks to multiple kernel functions.
Inline hook is called trampoline or detours hook, which is a method of receiving control when a hooked function is called. 
Dynamic kernel hook only requires loading a kernel module without kernel recompilation \newpar{or binary rewriting}, making itself easy to deploy in practice (e.g., Kprobes, Kpatch~\cite{kurmus2011attack,kurmus2014quantifiable,gebai2018survey}).

We leverage a library\footnote{https://github.com/cppcoffee/inl\_hook} to hook two kernel functions, i.e., \texttt{\_\_pte\_alloc} and \texttt{\_\_free\_pages}. 
\texttt{\_\_pte\_alloc} traces newly allocated L1PT pages. \texttt{\_\_free\_pages} monitors dynamically released pages. The collector hooks these two functions to update the three red-black trees as follows:

\vspace{2pt}\noindent$\bullet$ For a newly allocated L1PT page, its page, bank and row indexes will be updated into \texttt{pt\_rbtree} and  \texttt{pt\_row\_rbtree}, respectively. If there are new user pages that are adjacent to the L1PT page, they are added into \texttt{adj\_rbtree}. 

\vspace{2pt}\noindent$\bullet$ If an adjacent page is freed, it will be removed from \texttt{adj\_rbtree}. 

\vspace{2pt}\noindent$\bullet$ If an L1PT page is freed, it will be removed from \texttt{pt\_rbtree}. 
Also, the collector acquires a node in \texttt{pt\_row\_rbtree} that has the freed page's row index. 
Within the node, \texttt{pt\_count} in each \texttt{bank\_struct} corresponding to the freed page is decremented by one. If every \texttt{pt\_count} for the node becomes 0, then the node is deleted from \texttt{pt\_row\_rbtree}. 
Besides, the freed page's adjacent pages in \texttt{adj\_rbtree} are removed.

\begin{figure}
\centering
\includegraphics[width=\columnwidth]{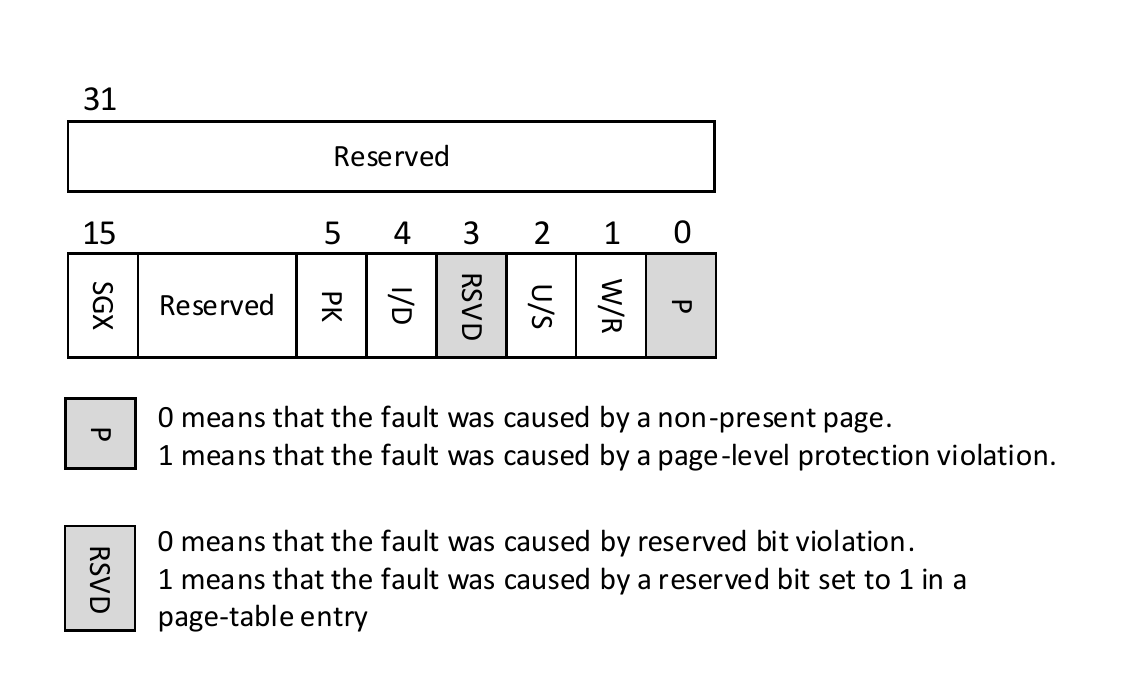}
\caption{Page-Fault Error Code.} 
\label{fig:page_fault}
\end{figure}

\subsection{Adjacent Page Tracer}\label{sec:tracer}
To trace memory accesses to adjacent pages at runtime, the adjacent page tracer leverages page fault handler.

\mypara{Page Fault Handler}
A page fault is a type of hardware exception. Whenever a user access to a virtual page violates access permissions dictated by one PTE, a page fault arises and will be captured by the MMU. As a response, the MMU will switch the process context to the kernel, which invokes the page fault handler to handle the fault based on an error code. The error code is generated by hardware and there are 7 page-fault error codes 
as shown in \autoref{fig:page_fault} (i.e., bits 0--5 and bit 15). 
For instance, when a memory access to a virtual address that is marked as non-present in the PTE (i.e., \texttt{present} bit is cleared), the access triggers a non-present page fault with \texttt{P} bit in the error code set to 0. To handle this page fault, the page fault handler can allocate a new physical page for the virtual address and marks the address as present in the PTE, the so-called \emph{demand paging}.

\mypara{Leverage Page Fault}
The adjacent page tracer can trace the memory access to a page by configuring flag bits in a PTE and hooking the page fault handler (i.e., \texttt{do\_page\_fault} function in the kernel space). 
As the memory access can be \emph{read}, \emph{write} or \emph{instruction fetch}, 
not every flag bit can be leveraged. For instance, a physical page becomes read-only when its corresponding PTE has \texttt{RW} bit cleared. Once write-access to the page occurs, a page fault is generated with \texttt{W/R} bit of the error code set to 1. As such, we experimented with each flag bit, results of which show that both \texttt{present} bit and \texttt{rsrv} bit in a PTE can be used for the tracing purpose. 
Next, we discuss why the tracer chooses \texttt{rsrv} bit rather than \texttt{present} bit.

Particularly, configuring \texttt{present} bit to trace the memory access causes a kernel crash, as the kernel performs active checks of \texttt{present} bit in a \new{leaf} PTE in multiple cases.
For instance, when a process is forking a new child process, the kernel checks \texttt{present} bit in the process's \new{leaf} PTEs. If 
one of the PTEs points to a physical page that is traced, \texttt{present} bit in the PTE is set to 0 by the tracer. When such a case occurs to the kernel check, the kernel will abort, because the tracer is unaware of when the forking occurs and it cannot restore \texttt{present} bit to 1 to pass the kernel check.

On top of that, we observe that one PTE has multiple \texttt{rsrv} bits in x86 which are unused and set to 0 by default. 
An access to a page with one \texttt{rsrv} bit in the PTE set to 1 will trigger a page fault and generate an error code of \texttt{RSVD} bit set to 1 shown in \autoref{fig:page_fault} 
\new{(this \texttt{RSVD} error has been leveraged in prior works ~\cite{basu2013efficient,bhattacharjee2013large,gandhi2014badgertrap,agarwal2017thermostat,wang2019safehidden} for different purposes)}. 
In contrast to the \texttt{present} bit check, the kernel does not check against \new{leaf} PTEs' \texttt{rsrv} bits. 
For instance, if an adjacent page is a part of a huge page of 2\,MiB, its \new{leaf} PTE is an L2PT entry and the kernel does not inspect any \texttt{rsrv} bit in the entry.
{As the page table management is a core component of the kernel, its code logic remains relatively stable. 
{Take a recent stable Linux kernel version (i.e., 5.10.4) as an example, there is no check against any \texttt{rsrv} bit, either. It is probably because that \texttt{rsrv} bits remain unused in \new{leaf} PTEs.
In our implementation, the tracer chooses a \texttt{rsrv} bit, i.e., bit 51 in the PTE.}}

\begin{figure}
\centering
\includegraphics[width=\columnwidth]{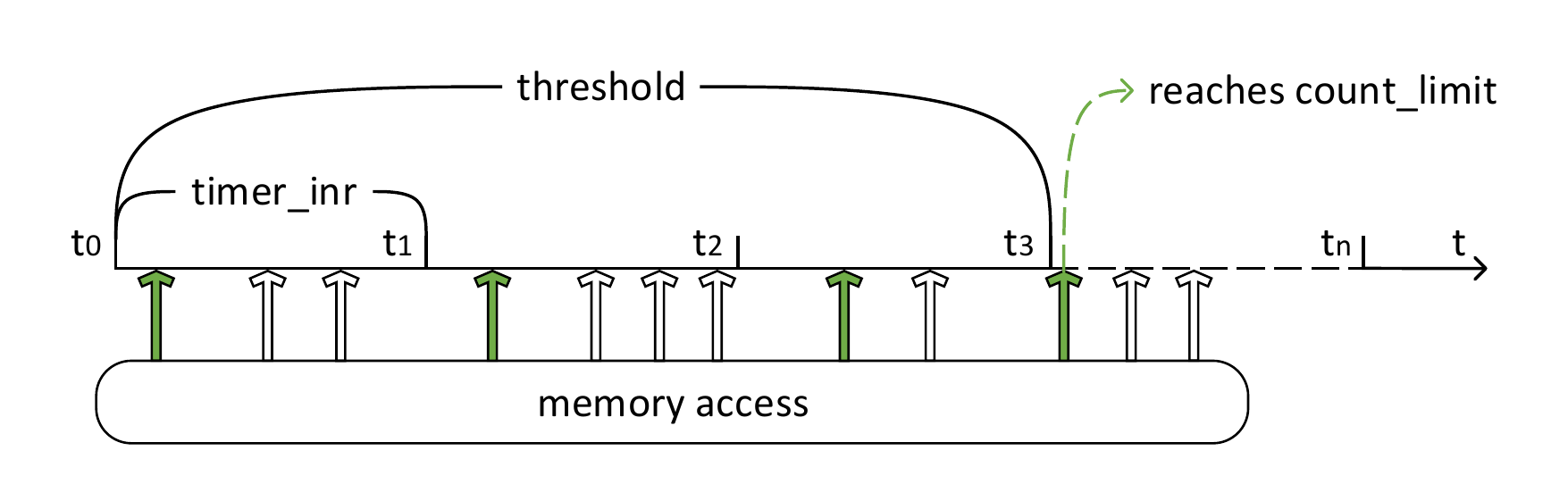}
\caption{The adjacent page tracer sets up tracing to adjacent pages in each time point from $t_0$, $t_1$, $t_2$, $t_3$, ..., $t_n$ and the interval between two adjacent time points is \texttt{timer\_inr}. The tracer captures the first memory access (highlighted in green) and ignores subsequent memory accesses in each interval of \texttt{timer\_inr} and updates \texttt{leak\_count}. Whenever \texttt{leak\_count} reaches \texttt{count\_limit}, the row refresher starts.} 
\label{fig:offline_profile}
\end{figure}

\mypara{Trace Adjacent Page}
Upon the tracer has configured \texttt{rsrv} bits in relevant PTEs pointing to the adjacent pages or the huge pages containing the adjacent pages, and flushed desired TLB entries, subsequent access to an adjacent page or its huge page will trigger a page fault. As \texttt{do\_page\_fault} is hooked, the tracer captures a faulting (huge) page with an expected error code of \texttt{RSVD} and collects complete DRAM information from the faulting (huge) page.
Thus, the tracer updates \texttt{leak\_count} of L1PT pages that are adjacent to either the captured (huge) page or its \new{leaf} page-table page. As an L1PT page may have multiple \texttt{bank\_struct}, \texttt{leak\_count} of each \texttt{bank\_struct} for the L1PT page should be updated accordingly.
If the \texttt{leak\_count} reaches a pre-determined limit in \autoref{fig:offline_profile}, row refresher will be triggered (see \autoref{sec:row_refresher}).



We note that the tracer \new{clears} \texttt{rsrv} bit before transferring control back to the user space to resume the memory access. However, any subsequent access to the same adjacent page or its huge page is no longer traced as \texttt{rsrv} bit is cleared. To address this issue, 
the tracer sets up a periodic timer to configure \texttt{rsrv} bit in a fixed interval and thus traces the accesses as frequently as possible. 
Specifically, when a timer comes, the tracer leverages kernel's \emph{reverse mapping} feature to translate a PPN in \texttt{adj\_rbtree} to a set of virtual addresses, as a PPN can be mapped to multiple virtual addresses. For each address, the tracer performs page-table walk, sets \texttt{rsrv} bit in its leaf PTE and flushes its cached TLB entry. 

It is clearly inefficient to do the reverse-mapping and page-table walk for every PPN in \texttt{adj\_rbtree} in every timer. To improve the efficiency, the tracer sets \texttt{rsrv} bit in PTEs relevant to the pages in \texttt{adj\_rbtree} and then frees corresponding nodes in \texttt{adj\_rbtree} in the first timer. If page faults with the error code of \texttt{RSVD} occur, the tracer captures them and stores the faulting addresses' PTE information into a dedicated ring buffer (i.e., \texttt{pte\_ringbuf}). When subsequent timers come, the tracer sets \texttt{rsrv} bits in PTEs stored in \texttt{pte\_ringbuf}, and handles remaining nodes in \texttt{adj\_rbtree} which are updated by the page table collector.

For any new page that is allocated for the user space in the default page fault handler, the tracer checks if its PPN or its L1PT page's PPN (if exists) is adjacent to any PPN in \texttt{pt\_rbtree}. If so, its \new{leaf} PTE information is inserted into \texttt{pte\_ringbuf}.

Particularly, \texttt{pte\_ringbuf} maintains two pointers for updates, i.e., \texttt{head} and \texttt{tail}.
If a new PTE is inserted to \texttt{pte\_ringbuf}, the \texttt{head} pointer is updated and points to the empty node next to the node of latest inserted PTE.
If one PTE is removed from \texttt{pte\_ringbuf} (i.e., its \texttt{rsrv} bit has been configured), the \texttt{tail} pointer is updated and points to the least recently inserted PTE. 
When the \texttt{head} and the \texttt{tail} point to the same ring buffer node, the buffer becomes empty.
The ring buffer size is pre-determined empirically. When the node number between the \texttt{tail} and the \texttt{head} pointers is no less than 80\% of the total node number of the ring buffer, the tracer allocates a larger ring buffer (e.g., four times of the old ring buffer size in our implementation), which will store newly inserted PTE. The old ring buffer will be freed when its stored PTEs are all consumed by the tracer.

As shown in \autoref{fig:offline_profile}, the time interval between two consecutive timers (denoted as $timer\_inr$) should be small enough to keep adjacent pages under close surveillance and \texttt{leak\_count} is updated promptly. 
On the other hand, our system might experience unacceptable overhead if the timer is too frequent and causes numerous context switches between user and kernel.
To this end, we discuss how to decide $timer\_inr$ in \autoref{sec:threshold} to keep \name's security guarantee while minimize its performance impacts.

\eat{
\begin{figure*}[ht]
	\centering
	\begin{minipage}[t]{\columnwidth}
		\centering
	    \includegraphics[width=\columnwidth]{image/refresh.pdf}
		\caption{Minimal \#ACTs per aggressor row required to induce bit flips in different BIOS refresh periods. The symbol ($\infty$) in a refresh period of 8ms means that no bit flip is observed using a large number of 10,000\,K ACTs.}
		\label{fig:refresh}
	\end{minipage}
	\hfill
	\begin{minipage}[t]{\columnwidth}
	    \centering
        \includegraphics[width=\columnwidth]{image/alg_cost.pdf}
        \caption{\name-induced costs are measured at different $timer\_inr$ in a real-world use case (i.e., Linux, Apache, MySQL and PHP (LAMP)-based server) and the costs are normalized to the cost when $timer\_inr$ is 1ms, showing that \name incurs the lowest overhead when $timer\_inr$ is 10ms.}
        \label{fig:alg_cost}
	\end{minipage}%
\end{figure*}
}

\subsection{Row Refresher}\label{sec:row_refresher}
\mypara{Direct-physical Map}
Linux systems and paravirtualized hypervisors (e.g., Xen) map the whole available physical memory directly into the kernel space~\cite{kernelmap,xenmap} in order for the kernel to access any data or code in the physical memory. Thus, every physical page allocated for the user space has been mapped to at least two virtual pages, i.e.,  a user virtual page and a kernel virtual page. While for a kernel's physical page, it is mapped to a single kernel virtual page. 

\mypara{Refresh Desired Rows}
If \texttt{leak\_count} in \texttt{bank\_struct} reaches a pre-determined limit (denoted as $count\_limit$), the row refresher refreshes desired rows specified by relevant \texttt{bank\_struct}.
As each node in \texttt{pt\_row\_rbtree} provides bank indexes and row indexes, the refresher leverages them to reconstruct a physical address. Based on the {direct-physical map}, the refresher finds out a kernel virtual address mapped to the physical address. 
As a read-access to a row can automatically re-charge the row and prevent potential bit flips, the refresher flushes CPU caches of the kernel virtual address, reads the virtual address, and resets \texttt{leak\_count} to 0 at last.

If $count\_limit$ is set too small (e.g., 1), the refreshing cost may become unacceptable as many unnecessary refreshes are introduced by regular memory accesses to adjacent pages. If $count\_limit$ is too large, the refresher is unable to promptly refresh a row before it is flipped. Thus, $count\_limit$ should be no less than 2 and we decide its value in the next section. 

\subsection{Offline Profile}\label{sec:threshold}
\name decides realistic and reasonable $timer\_inr$ and $count\_limit$ to keep its security and practicality design principles. As illustrated in \autoref{fig:offline_profile}, \new{the adjacent page tracer only captures the first memory access to an adjacent page within each $timer\_inr$ and updates \texttt{leak\_count}. The subsequent memory accesses within $timer\_inr$ to the same page will be ignored by the tracer. Thus, the maximum time period (denoted as $threshold$) for hammer before the page is refreshed has such an equation: ${threshold = timer\_inr \times (count\_limit - 1)}$.
This means that \name must carefully set $threshold$ short enough to ensure that no bit flip occurs within $threshold$.}

\eat{
\begin{figure}
\centering
\includegraphics[width=\columnwidth]{image/refresh.pdf}
\caption{Minimal \#ACTs per aggressor row required to induce bit flips in different BIOS refresh periods for DDR3 and DDR4-based computing systems. The symbol ($\infty$) in a refresh period of 8ms means that no bit flip is observed using a large number of 10,000\,K ACTs.}
\label{fig:refresh}
\end{figure}
}

We decide $threshold$ based on the equation: ${threshold = \texttt{tRC} \times \#ACT}$, where \texttt{tRC} is the time interval between two successive ${ACT}$ commands and $\#ACT$ is the number of activations for all the hammered rows that is required to induce the first bit flip. Thus, we guarantee that no bit flip occurs within the time interval of $threshold$.
We learn from Kim et al.~\cite{kim2020revisiting} that \texttt{tRC} is around 50\,ns and $\#ACT$ is in the order of 20\,K on DDR3 modules and 10\,K on DDR4 modules. Compared to DDR3 modules that require at least 1 aggressor row, no less than 2 aggressor rows are required in DDR4 modules due to the ChipTRR. As such, $\#ACT$ for triggering the first bit flip is around 20\,K for both DDR3 and DDR4 modules.
To this end, $threshold$ is set to 1\,ms, below which DRAM modules are believed to be rowhammer-free. 
As both $timer\_inr$ and $count\_limit$ for \name are unsigned integers, $timer\_inr$ is set to 1\,ms and $count\_limit$ is set to 2.

\section{Security Evaluation}\label{sec:eva}
We now turn to evaluate the security effectiveness of \name on three different hardware configurations, summarized in \autoref{tab:config}, all running Ubuntu.

We deploy \name into each system against one representative kernel privilege escalation attack, i.e., Memory Spray~\cite{seaborn2015exploiting} that hammers user memory adjacent to L1PTEs, CATTmew~\cite{cheng2018still} that hammers device driver buffer adjacent to L1PTEs, and PThammer~\cite{zhang2019telehammer} that implicitly hammers L1PTEs adjacent to other L1PTEs. Both Memory Spray and CATTmew are explicit rowhammer attacks with two different types of memory accessible to unprivileged users. PThammer is the only implicit rowhammer attack. 


\begin{table*}
\centering
\begin{tabular}{lcccccc}
\toprule
\multicolumn{1}{l}{\multirow{2}{*}{\textbf{Machine Model}}} & 
\multicolumn{3}{c}{\textbf{Hardware Configuration}} &
\multirow{1}{*}{\textbf{Attack}} &
\multirow{1}{*}{\textbf{\name}} \\ \cline{2-4}
 & 
 \multirow{1}{*}{{CPU Arch.}} & \multirow{1}{*}{{CPU Model}} & \multirow{1}{*}{{DRAM (Part No.)}} & $m$ Targeted Victim Pages
 & 
 \multicolumn{1}{c}{Bit Flip Failed?} \\\hline
 \multirow{2}{*}{{Dell Optiplex 390}} & \multirow{2}{*}{KabyLake} & \multirow{2}{*}{i7-7700k} & \multirow{1}{*}{Kingston DDR4} & \multirow{2}{*}{Memory Spray~\cite{seaborn2015exploiting}} & \multirow{2}{*}\CheckmarkBold \\
  &  &  &  (99P5701-005.A00G) &  & \\ \hline
  \multirow{2}{*}{{Dell Optiplex 990}} & \multirow{2}{*}{SandyBridge} & \multirow{2}{*}{i5-2400} & {Samsung DDR3} & \multirow{2}{*}{CATTmew~\cite{cheng2018still}} & \multirow{2}{*}\CheckmarkBold \\
  
  &  &  &  (M378B5273DH0-CH9) &  & \\ \hline
 \multirow{2}{*}{{Thinkpad X230}} & \multirow{2}{*}{IvyBridge} & \multirow{2}{*}{i5-3230M} & {Samsung DDR3} & \multirow{2}{*}{PThammer~\cite{zhang2019telehammer}} & \multirow{2}{*}\CheckmarkBold \\
  &  &  &  (M471B5273DH0-CH9) &  & \\
\bottomrule
\end{tabular}
\caption{Each rowhammer attack targets $m$ (e.g., 50 in our experiments) victim pages of L1PTEs. With \name enabled, each attack fails to induce bit flips in these pages, indicating that those attacks have been mitigated.}
\label{tab:config}
\end{table*}

\subsection{Defeat Memory Spray}
\mypara{Background}
The Memory Spray~\cite{seaborn2015exploiting} is the first rowhammer attack targeting L1PTs. It is a probabilistic attack, as it sprays numerous L1PT pages into the memory with the hope that some L1PT pages are placed onto victim rows adjacent to attacker-controlled rows. As such, exploitable bits in L1PTEs can be flipped, resulting in kernel privilege escalation. 

\mypara{Evaluation Details}
We test the effectiveness of \name against the Memory Spray on the Dell Optiplex 390.
In this machine, traditional 2-sided hammer pattern cannot trigger any bit flip and instead we use the 3-sided hammer identified by TRRespass.
We first conduct 3-sided hammer to randomly identify $m$ (e.g., 50 in our evaluation) vulnerable pages that have reproducible bit flips, that is, a vulnerable page has at least one victim physical address ($P_v$) and hammering three aggressor addresses $P_a$, $P_b$ and $P_c$ will flip bits in $P_v$. 

We then optimize the attack by using the kernel privilege to put page tables onto vulnerable pages in a deterministic way. Specifically, we spray $m$ pages of L1PTs by creating a virtual memory region of $2m$\,MiB, ask the kernel to 
copy the content of the $m$ pages of L1PTs into the $m$ vulnerable pages, which are then used to translate the virtual memory region. The vulnerable pages now contain L1PTs and the original L1PTs are removed.
By doing so, an attacker will definitely corrupt any one of the L1PTs pages by hammering three relevant aggressor addresses. 
When \name is enabled to collect and protect the $m$ pages of L1PTs, we re-start the optimized attack for $m$ hours (one-hour hammer for one vulnerable L1PT page) and observe no single bit flip in those $m$ pages of L1PTs by checking their integrity, indicating that the Memory Spray attack has been successfully defeated.

\subsection{Defeat CATTmew}\label{sec:cattmew}
\mypara{Background}
As mentioned in \autoref{sec:bkgd}, CATT~\cite{brasser17can} enforces physical user-kernel isolation. 
CATTmew~\cite{cheng2018still} breaks CATT's security guarantee by identifying device (e.g., SCSI Generic)  driver buffers that are kernel memory but can be accessed by unprivileged users. CATTmew exploits the driver buffers to ambush adjacent L1PT pages for hammer, with the hope that these L1PT pages are prone to bit flips. 

\mypara{Evaluation Details}
We use 2-sided hammer to search $m$ vulnerable pages on the Dell Optiplex 990.
A vulnerable page has at least one victim physical address ($P_v$) and hammering two aggressor addresses ($P_a$ and $P_b$) flips bits in $P_v$. 

We then rely on the kernel privilege to convert CATTmew into a deterministic attack. Specifically, we spray $m$ L1PT pages and copy their entries onto the $m$ vulnerable pages as what we did in the optimized Memory Spray attack. On top of that, we apply for the SCSI Generic (SG) buffer using Linux user APIs. In this test machine, we can apply as large as $123$\,MiB and only $8m$\,KiB of the SG buffer \new{are} enough. We instruct the kernel to copy the allocated SG buffer's content into the $2m$ aggressor pages and change the buffer's address mappings accordingly. To this end, hammering the buffer will induce bit flips in the vulnerable L1PT pages. 
However, when \name is set active, no single bit flip has been observed in those L1PT pages after $m$ hours of hammering, indicating that \name is effective in defeating the CATTmew attack. 

\subsection{Defeat PThammer}
\label{sec:pthammer}
\mypara{Background}
Rowhammer attacks before PThammer~\cite{zhang2019telehammer} are explicit rowhammer that require access to an exploitable aggressor row (e.g. adjacent to a row of L1PTs). PThammer {voids} this requirement. 
By spaying L1PT pages and placing some onto victim rows with a high probability, PThammer exploits page-table walk to produce frequent loads of some L1PTEs from aggressor rows (i.e., ``implicitly hammering L1PTEs"), which will induce bit flips in other L1PTEs in victim rows.

\mypara{Evaluation Details}
We optimize PThammer by using the kernel privilege to present a more efficient and deterministic attack on the Thinkpad X230. 
Specifically, PThammer uses eviction sets to flush TLB entries and CPU caches of desired L1PTEs and user memory loads trigger the page-table walk to implicitly hammer the L1PTEs. However, the eviction-based flush is probabilistic. In our test, the kernel assists PThammer in performing the flush through explicit instructions (i.e., \texttt{invlpg} for TLB flush and \texttt{clflush} for L1PTEs flush). 
Thus, its hammer instruction sequence is kernel-assisted flush with a user memory load, which is less efficient than the aforementioned 2-sided hammer that applies \texttt{clflush} for user data flush. In such a case, we cannot use the traditional 2-sided hammer to identify vulnerable pages, as these pages may become non-flippable to the kernel-assisted hammer. 
To address this issue, we add a certain number of \texttt{NOP} (e.g., 180) instructions into the 2-sided hammer instruction sequence to meet the time cost taken by the kernel-assisted hammer. By doing so, $m$ vulnerable pages of interest can be discovered. 

As PThammer massages L1PTEs onto vulnerable pages with a probability, we instead spray $3m$ number of L1PT pages by creating a virtual memory region of $6m$\,MiB. We then ask the kernel to copy all entries of the L1PT pages into the $m$ vulnerable pages and the $2m$ aggressor pages. The kernel then changes the address mappings of the created virtual memory region by using the new $3m$ L1PT pages. As such, the optimized PThammer successfully induces bit flips in the $m$ vulnerable L1PT pages by using the kernel-assisted hammer against the $2m$ aggressor L1PT pages.
In comparison, we enable \name before starting the optimized PThammer. As each $2$ aggressor L1PT pages is adjacent to a vulnerable L1PT page in \texttt{pt\_rbtree}, \name traces memory accesses to the created virtual pages pointed by the L1PT page entries. Considering that the PThammer still requires frequent memory loads of the created virtual pages for page-table walk, it cannot bypass the tracing. 
After $m$ hours of hammering the $2m$ aggressor pages, no bit flip occurs, meaning that \name has mitigated PThammer. 

\section{Performance Evaluation}\label{sec:perf_eva}

We evaluate the performance impacts including memory consumption induced by \name. The experiments are conducted in a DDR4-based system. The system is Ubuntu running on top of a Dell Desktop with Intel i7-7700K and Samsung 16\,GiB DDR4 (part number: M378A2G43AB3-CWE). 
By default, the row distance implemented by \name between adjacent rows and L1PT-page rows is up to 6-row, denoted by $\Delta_{\pm6}$. In comparison, we also measure its impacts in the scenario of only one-row-distance that previous works (e.g., ~\cite{konoth2018zebram}) assume, denoted by $\Delta_{\pm1}$.
The results show that \name in both scenarios of  $\Delta_{\pm6}$ and $\Delta_{\pm1}$ incurs an average slow-down \new{within 0.83\%}
indicating that the row distance may have a relatively small impact on the performance overhead. \new{We note that the cost of initially loading \name into the kernel is around 28\,ms and it occurs only once.}
We also validate the system robustness of \name, results of which show that \name does not affect the stability of the protected system, making itself practical.

\subsection{Benchmark Runtime Overhead}~\label{sec:benchmarks}
We measure \name-induced runtime overhead using two popular benchmarks, i.e., \texttt{SPECspeed 2017 Integer}~\cite{specint2017} and \texttt{Phoronix} test suite\footnote{https://github.com/phoronix-test-suite/phoronix-test-suite}. 

\new{
\texttt{SPEC CPU 2017} is an industry standard benchmark package that contains CPU-intensive programs for measuring compute-intensive performance. It has 43 benchmarks in total and is organized into 4 suites, among which \texttt{SPECspeed} 2017 Integer has been used. 
This suite launches $10$ integer programs with a specific configuration file customized from \emph{Example-linux-gcc-x86.cfg} and the benchmark results are summarized in \autoref{tab:spec}.
As we can see from the table, the overhead of $\Delta_{\pm6}$ (0.83\%) and $\Delta_{\pm1}$ (0.07\%) are less than 1\%.
}


\texttt{Phoronix} is a free and open-source benchmark software for mainstream OSes (e.g., Linux, MacOS and Windows). It allows for testing performance overhead against common applications in an automated manner. 
As this suite has a large number of programs testing different aspects of a system, we select a subset of the available programs to stress-test performance of CPU, memory, network I/O and disk I/O.
As shown in \autoref{tab:phoronix}, the average performance overhead is 0.22\% for $\Delta_{\pm1}$ and 0.24\% for $\Delta_{\pm6}$, respectively, indicating that the \texttt{Phoronix} overhead is negligible in both scenarios. 

\begin{table}
\centering
\begin{tabular}{lcc}
\toprule
\multirow{2}{*}{\textbf{Programs}} & \multicolumn{2}{c}{\textbf{\name Overhead}} \\\cline{2-3}
& \multirow{1}{*}{\textbf{$\Delta_{\pm1}$}} & {\textbf{$\Delta_{\pm6}$} (default)}\\ \hline 
{perlbench\_s}   & {0.67\%} & 0.67\% \\
{gcc\_s}  & {0.23\%} & 0.92\% \\ 
{mcf\_s}  & {-0.76\%} & 0.30\% \\
{omnetpp\_s}   & {-0.81\%} & 1.82\%\\
{xalancbmk\_s}  & {0.36\%} & 2.50\% \\
{x264\_s}  & {0.00\%} & 0.61\% \\
{deepsjeng\_s} &  {0.00\%} & 0.28\%\\
{leela\_s}   & {0.23\%} & 0.46\% \\
{exchange2\_s}   & {-0.70\%} & -0.23\% \\
{xz\_s}   & {1.48\%} & 0.93\% \\ \hline
{\textbf{Mean}}  & {0.07\%} & 0.83\% \\
\bottomrule
\end{tabular}
\caption{\new{\texttt{SPECspeed} 2017 Integer benchmark results.}}
\label{tab:spec}
\end{table}

\begin{table}
\centering
\begin{tabular}{lcc}
\toprule
\multirow{2}{*}{\textbf{Programs}} & \multicolumn{2}{c}{\textbf{\name Overhead}} \\\cline{2-3}
& \multirow{1}{*}{\textbf{$\Delta_{\pm1}$}} & {\textbf{$\Delta_{\pm6}$}}\\ \hline 
{Apache} & {-0.16\%} & {0.32\%} \\
{unpack-linux} & {1.31\%} & {1.84\%} \\
{iozone} & {0.89\%} & {-1.15\%}\\
{postmark} & {0.89\%} & {0.00\%} \\
{stream:Copy} &  {0.01\%} & {0.00\%} \\
{stream:Scale} & {0.6\%}  & {0.23\%} \\
{stream:Triad} & {0.07\%} & {0.37\%} \\
{stream:Add} &   {0.03\%} & {0.35\%} \\
{compress-7zip} & {1.52\%} & {2.24\%} \\
{openssl} & {0.14\%} & {0.13\%}   \\
{pybench} &  {0.00\%} & {0.52\%}   \\
{phpbench} &  {0.92\%} & {0.01\%} \\
{cacheben:read} & {-0.38\%} & {0.26\%}  \\
{cacheben:write} &   {-0.26\%} & {-0.44\%}   \\
{cacheben:modify} &   {-0.01\%} & {0.67\%}   \\
{ramspeed:INT} &   {-0.09\%} & {-0.63\%}   \\
{ramspeed:FP} &   {-0.15\%} & {-0.63\%} \\
\hline
{\textbf{Mean}} & {0.22\%} & {0.24\%} \\
\bottomrule
\end{tabular}
\caption{\texttt{Phoronix} benchmark results.}
\label{tab:phoronix}
\end{table}

\subsection{LAMP Runtime Memory Consumption}\label{sec:lamp}
We use a real-world use case to measure runtime memory consumption of \name, that is, a LAMP server (i.e., Linux, Apache, MySQL and PHP). 
We run a common tool (i.e., \texttt{Nikto}~\cite{Nikto}) in another machine for 60 minutes to stress test the LAMP server.
\texttt{Nikto} is a web server scanner that tests the LAMP server for insecure files and outdated server software. It also carries out generic and server type specific checks.

\begin{figure}
\centering
\includegraphics[width=\columnwidth]{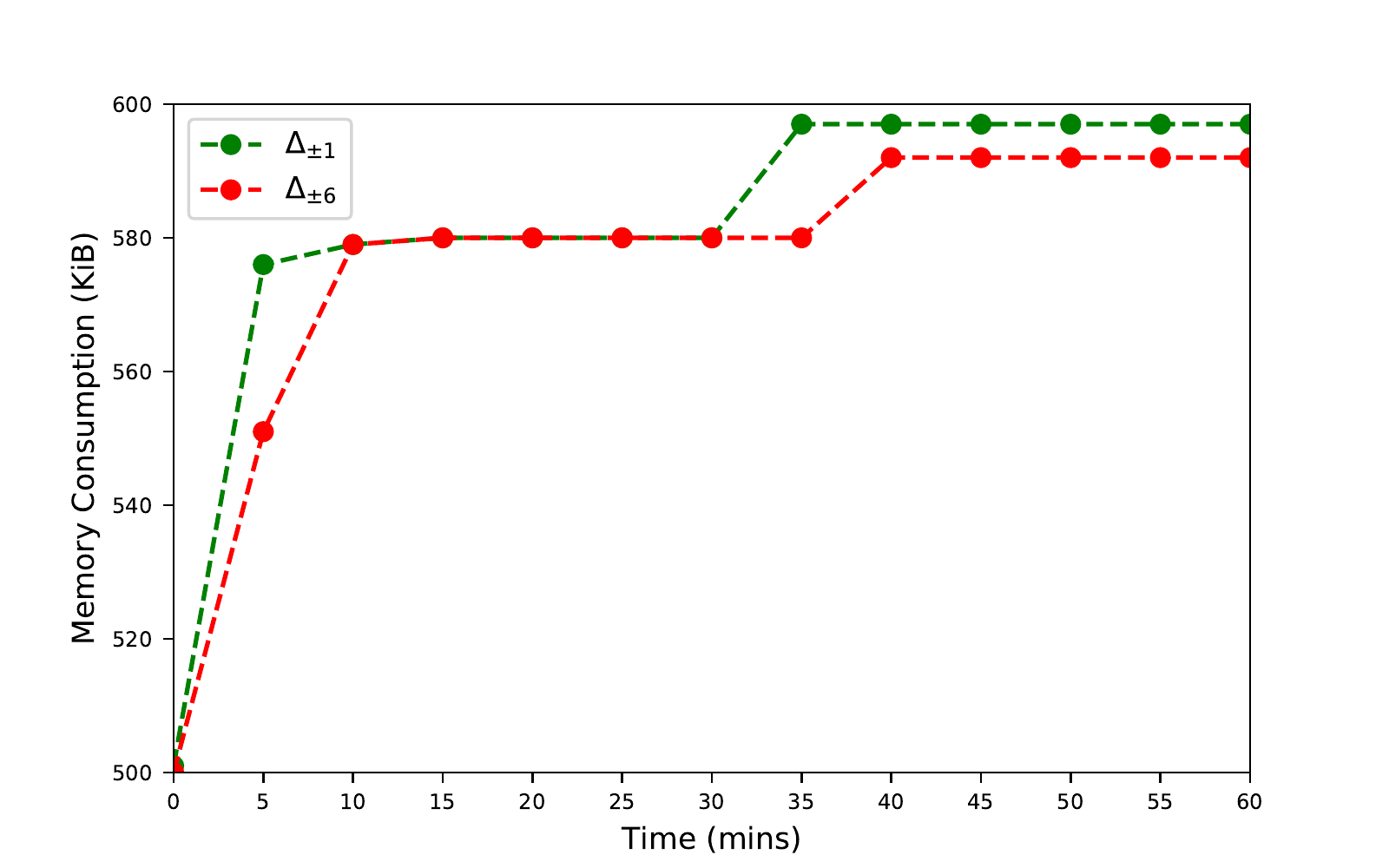}
\caption{The memory consumed by \name in both
$\Delta_{\pm1}$ and $\Delta_{\pm6}$ for the LAMP production environment.}
\label{fig:mem_cost}
\end{figure}

The memory cost induced by \name within the 60 minutes is shown in \autoref{fig:mem_cost}. 
The memory consumption is a total memory size of three red-black trees (i.e., \texttt{pt\_rbtree}, \texttt{pt\_row\_rbtree} and \texttt{adj\_rbtree}) and the ring buffer (i.e., \texttt{pte\_ringbuf}). We note that 
the pre-allocated \texttt{pte\_ringbuf} is 396\,KiB.
As shown in the figure, the memory costs in both $\Delta_{\pm1}$ and $\Delta_{\pm6}$ increase gradually and reach a relatively stable level in the last 15 minutes. Both $\Delta_{\pm1}$ and $\Delta_{\pm6}$ have a similar and low memory cost (i.e., less than 600\,KiB).

\begin{figure}
\centering
\includegraphics[width=\columnwidth]{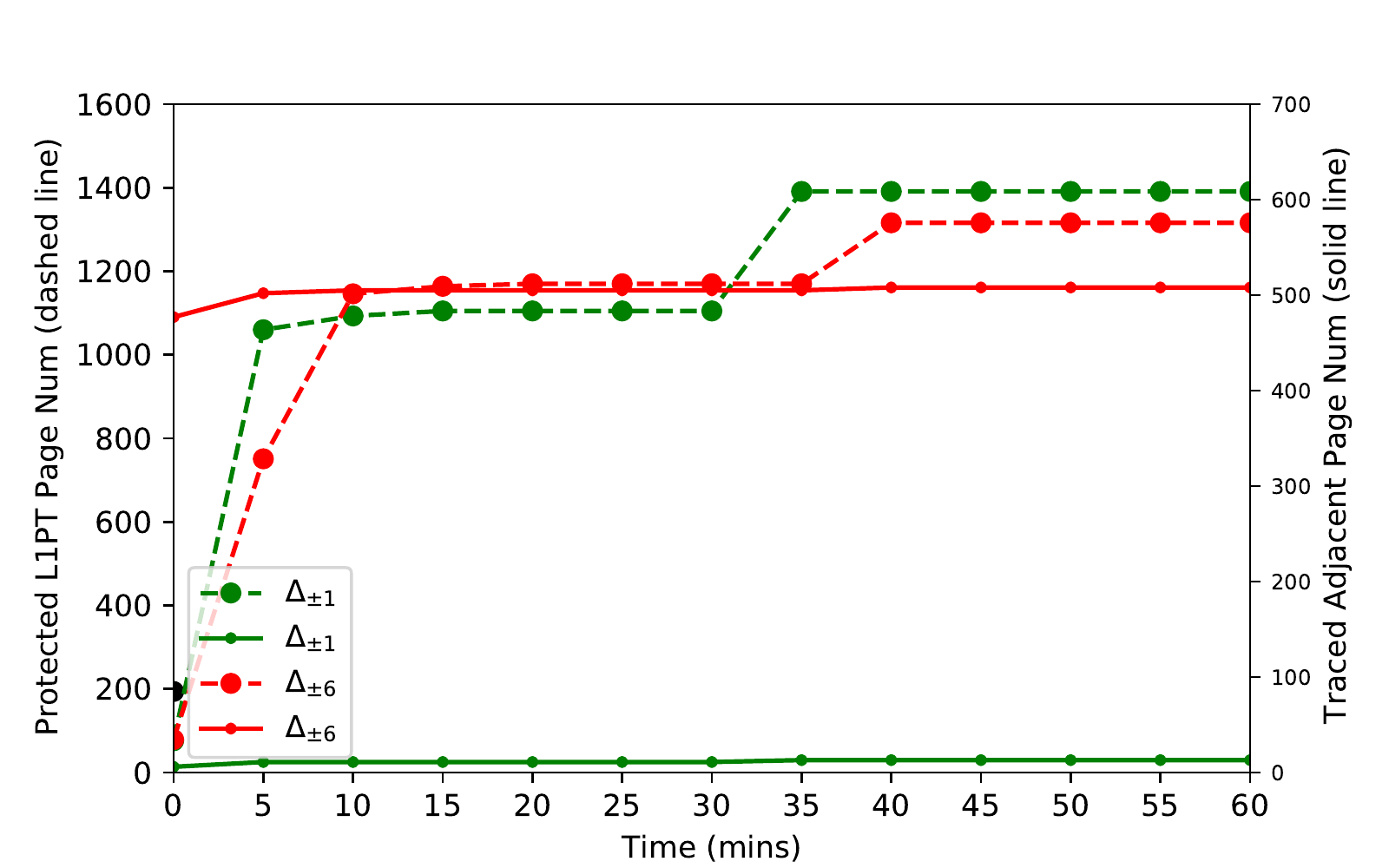}
\caption{The numbers of protected L1PT pages and traced adjacent pages in both
$\Delta_{\pm1}$ and $\Delta_{\pm6}$ for the LAMP production environment.}
\label{fig:node_cost}
\end{figure}
\mypara{Protected and Traced Page Number}
When computing the memory consumption, we also collect the unique page numbers that \name protects and traces, respectively. \autoref{fig:node_cost} shows that both protected L1PT page number and traced adjacent page number in $\Delta_{\pm1}$ and $\Delta_{\pm6}$ increase and become stable within the 60 minutes.
We note that the protected L1PT page numbers in $\Delta_{\pm1}$ and $\Delta_{\pm6}$ are in the same order of magnitude as the overall system activities in both scenarios are similar to each other. 
As an L1PT-page row in $\Delta_{\pm6}$ can have up to 12 adjacent rows, 6 times the adjacent row number that an L1PT-page row can have in $\Delta_{\pm1}$, more adjacent pages are expected to be collected in $\Delta_{\pm6}$. 
\autoref{fig:node_cost} indeed shows that the traced adjacent page number in $\Delta_{\pm6}$ is higher than that in $\Delta_{\pm1}$. 

\begin{table}
\centering
\begin{tabular}{lllcc}
\toprule
\multicolumn{2}{l}{\multirow{3}{*}{\textbf{Linux Test Project}}} & \multirow{3}{*}{{\textbf{Vanilla System}}} & \multicolumn{2}{c}{\multirow{1}{*}{{\textbf{\name}}}} \\
 & & & $\Delta_{\pm1}$ & $\Delta_{\pm6}$ \\
  & & & & (default)\\
\hline
\multirow{4}{*}{\textbf{File}} & {open} & \multicolumn{1}{c}{\CheckmarkBold} & \multicolumn{1}{c}{\CheckmarkBold} & \multicolumn{1}{c}{\CheckmarkBold}\\
& {close} & \multicolumn{1}{c}{\CheckmarkBold} & \multicolumn{1}{c}{\CheckmarkBold} & \multicolumn{1}{c}{\CheckmarkBold}\\
& {ftruncate} & \multicolumn{1}{c}{\CheckmarkBold} & \multicolumn{1}{c}{\CheckmarkBold} & \multicolumn{1}{c}{\CheckmarkBold} \\
& {rename} & \multicolumn{1}{c}{\CheckmarkBold} & \multicolumn{1}{c}{\CheckmarkBold} & \multicolumn{1}{c}{\CheckmarkBold}\\ \hline
\multirow{4}{*}{\textbf{Network}} 
& {Listen} & \multicolumn{1}{c}{\CheckmarkBold} & \multicolumn{1}{c}{\CheckmarkBold} & \multicolumn{1}{c}{\CheckmarkBold}\\ 
& {Socket} & \multicolumn{1}{c}{\CheckmarkBold} & \multicolumn{1}{c}{\CheckmarkBold} & \multicolumn{1}{c}{\CheckmarkBold}\\
& {Send} & \multicolumn{1}{c}{\CheckmarkBold} & \multicolumn{1}{c}{\CheckmarkBold} & \multicolumn{1}{c}{\CheckmarkBold}\\
& {Recv} & \multicolumn{1}{c}{\CheckmarkBold} & \multicolumn{1}{c}{\CheckmarkBold} & \multicolumn{1}{c}{\CheckmarkBold}\\  \hline
\multirow{6}{*}{\textbf{Memory}} & {mmap} & \multicolumn{1}{c}{\CheckmarkBold} & \multicolumn{1}{c}{\CheckmarkBold} & \multicolumn{1}{c}{\CheckmarkBold}\\
& {munmap} & \multicolumn{1}{c}{\CheckmarkBold} & \multicolumn{1}{c}{\CheckmarkBold} & \multicolumn{1}{c}{\CheckmarkBold}\\
& {brk} & \multicolumn{1}{c}{\CheckmarkBold} & \multicolumn{1}{c}{\CheckmarkBold} & \multicolumn{1}{c}{\CheckmarkBold}\\
& {mlock} & \multicolumn{1}{c}{\CheckmarkBold} & \multicolumn{1}{c}{\CheckmarkBold} & \multicolumn{1}{c}{\CheckmarkBold}\\
& {munlock} & \multicolumn{1}{c}{\CheckmarkBold} & \multicolumn{1}{c}{\CheckmarkBold} & \multicolumn{1}{c}{\CheckmarkBold}\\
& {mremap} & \multicolumn{1}{c}{\CheckmarkBold} & \multicolumn{1}{c}{\CheckmarkBold} & \multicolumn{1}{c}{\CheckmarkBold}\\\hline
\multirow{3}{*}{\textbf{Process}} & {getpid} & \multicolumn{1}{c}{\CheckmarkBold} & \multicolumn{1}{c}{\CheckmarkBold} & \multicolumn{1}{c}{\CheckmarkBold}\\
& {exit} & \multicolumn{1}{c}{\CheckmarkBold} & \multicolumn{1}{c}{\CheckmarkBold} & \multicolumn{1}{c}{\CheckmarkBold}\\
& {clone} & \multicolumn{1}{c}{\CheckmarkBold} & \multicolumn{1}{c}{\CheckmarkBold} & \multicolumn{1}{c}{\CheckmarkBold}\\ \hline
\multirow{3}{*}{\textbf{Misc.}} & {ioctl} &\multicolumn{1}{c}{\CheckmarkBold} & \multicolumn{1}{c}{\CheckmarkBold} & \multicolumn{1}{c}{\CheckmarkBold}\\
& {prctl} & \multicolumn{1}{c}{\CheckmarkBold} & \multicolumn{1}{c}{\CheckmarkBold} & \multicolumn{1}{c}{\CheckmarkBold}\\
& {vhangup} & \multicolumn{1}{c}{\CheckmarkBold} & \multicolumn{1}{c}{\CheckmarkBold} & \multicolumn{1}{c}{\CheckmarkBold} \\ 
\bottomrule 
\end{tabular}
\caption{System-call stress tests from Linux Test Project.}
\label{tab:test}
\end{table}

\subsection{System Robustness}
To evaluate the robustness of our test system after deploying \name, we select 20 system calls of different types and perform stress tests for each selected system call on both the vanilla system and the \name-based system. The stress tests come from Linux Test Project (LTP)\footnote{https://github.com/linux-test-project/ltp} and they are used to identify system problems.
As can be seen from \autoref{tab:test}, the stress test results clearly show that there is no deviation for the \name-based system compared to the vanilla system. 
Also, we do not observe any issue when executing previous benchmarks. As a result, the test system runs \new{stably} with \name enabled.

\section{Discussion}\label{sec:dis}

\mypara{Root Privilege Escalation Attack}
Rowhammer root privilege escalation attack is that an unprivileged adversary gains root privilege by corrupting opcodes of a \emph{setuid} process~\cite{gruss2017another}.
This only known root privilege escalation attack on x86 has already been effectively and efficiently defeated
by RIP-RH~\cite{bock2019rip} that physically isolates sensitive user processes. 
In addition, \name can also be extended to defend against this attack. As described in Section~\ref{sec:overview}, \name treats page tables as protected objects. Thus, trusted user can pass specified objects (i.e., binary code pages of \texttt{setuid} processes) to \name through a provided user API and \name uses similar mechanisms to protect those objects.

\mypara{DMA-based Kernel Privilege Escalation Attack}
There is NO existing DMA-based kernel privilege escalation attack on x86. Such attack is demonstrated on ARM (Drammer~\cite{van2016drammer}), and it has been defeated by GuardION~\cite{van2018guardion} that enforces DMA memory isolation. 
In the future, if such attacks on x86 prove to be feasible, we can take the following two ways to solve. One is to integrate \name with existing orthogonal defenses. In particular, 
ALIS~\cite{tatar2018throwhammer} on x86 physically isolates DMA memory using guard rows and bit flips are thus confined to DMA memory of attackers.

Alternatively, \name can leverage IOMMU~\cite{intel} to monitor remote access to DMA memory by configuring I/O page tables, similar to MMU-based page tables. 
Specifically, \name collects (I/O) page tables and their adjacent DMA memory pages that are allocated to users. By configuring I/O page tables, \name traces accesses to the collected DMA pages.
When IOMMU is widely available on x86, we believe that \name can leverage it to defend (I/O) page tables against unknown DMA-based kernel privilege escalation attacks.

\new{\mypara{Support for ARM Architecture}
Although there are reserved bits in page table entries in the ARM architecture, setting these bits will not trigger any hardware fault~\cite{arm}. If we extend \name to provide ARM support, a possible solution is to disable the page table walk and capture the address-translation fault. However, this solution might introduce a larger performance overhead, as each memory access to a process triggers the fault if the process has pages adjacent to L1PT pages. 
Alternatively, we can leverage the present bit rather than the reserved bit in both x86 and ARM. As discussed in Section~\ref{sec:tracer}, the kernel performs active checks of the present bit in a leaf PTE. To address this issue, we can leverage the approach~\cite{wang2019safehidden} to find all the functions where the kernel performs the check. By hooking these functions, we can restore the present bit and bypass the kernel check.
}

\mypara{Level-1 and Higher-level Page Table}
Existing kernel privilege escalation attacks focus on \new{corrupting} L1PTs, and there is no demonstrated attack that has successfully exploited higher-level page tables~\cite{wu2018CAT}. 
If such an attack may be feasible in the future, we can easily extend our \name to protect higher-level page tables. For instance, when \name is extended to protect L2PT pages , \name collects desired user pages if they or their corresponding L1PT or L2PT pages are adjacent to either L1PT or L2PT pages. \name traces the collected user pages by setting \texttt{rsrv} bits in their \new{leaf} PTEs and refreshes relevant page-table pages when necessary. As the number of higher-level PT pages is significantly smaller than the number of L1PT pages (e.g., an L2PT page can point up to 512 L1PT pages), we believe that the additional performance overhead will not be high.

\eat{
\mypara{Hammer Sensitive Objects Implicitly}
Besides PThammer~\cite{zhang2019telehammer} that implicitly hammers page-table through page-table walk, there are also concerns about whether other implicit rowhammer attacks exist. 
To the best of our knowledge, there is only one possible candidate but it has not demonstrated a successful attack. 
Zhang et al.~\cite{zhang2020ghostknight} have shown how to leverage an optimization feature of modern processors (i.e., speculative execution) to mount a rowhammer trial but only flip bits in attacker-accessible objects.  

}


\section{Conclusion}\label{sec:conclusion}
In this paper, we proposed a software-only defense, named \name, that
protects level-1 page tables against rowhammer attacks on x86. 
\name is a loadable kernel module and compatible with commodity Linux systems without requiring any kernel modification. 

We evaluated the security effectiveness of \name-enabled systems using three kernel privilege escalation attacks. 
Also, we measured \name's performance overhead, memory cost, and stability using multiple benchmark suites and a real-world use case.
The experimental results indicate that \name is effective in defending against all the mentioned attacks, and practical in incurring low performance overhead and memory cost. Besides, it does not affect the system stability. 

\bibliographystyle{plain}
\bibliography{defs,main}

\begin{thebibliography}{10}

\bibitem{agarwal2017thermostat}
Neha Agarwal and Thomas~F Wenisch.
\newblock Thermostat: Application-transparent page management for two-tiered
  main memory.
\newblock In {\em Architectural Support for Programming Languages and Operating
  Systems}, pages 631--644, 2017.

\bibitem{Apple}
{Apple, Inc.}
\newblock About the security content of mac efi security update 2015-001.
\newblock \url{https://support.apple.com/en-au/HT204934}, August 2015.

\bibitem{arm}
{ARM, Inc.}
\newblock Arm architecture reference manual armv8, for armv8-a architecture
  profile.
\newblock \url{https://developer.arm.com/documentation/ddi0487/gb}.

\bibitem{aweke2016anvil}
Zelalem~Birhanu Aweke, Salessawi~Ferede Yitbarek, Rui Qiao, Reetuparna Das,
  Matthew Hicks, Yossi Oren, and Todd Austin.
\newblock {ANVIL}: Software-based protection against next-generation rowhammer
  attacks.
\newblock In {\em Architectural Support for Programming Languages and Operating
  Systems}, pages 743--755, 2016.

\bibitem{basu2013efficient}
Arkaprava Basu, Jayneel Gandhi, Jichuan Chang, Mark~D Hill, and Michael~M
  Swift.
\newblock Efficient virtual memory for big memory servers.
\newblock In {\em International Symposium on Computer Architecture}, pages
  237--248, 2013.

\bibitem{bennettpanopticon}
Tanj Bennett, Stefan Saroiu, Alec Wolman, and Lucian Cojocar.
\newblock Panopticon: A complete in-dram rowhammer mitigation.
\newblock In {\em Workshop on DRAM Security}, 2021.

\bibitem{bhattacharjee2013large}
Abhishek Bhattacharjee.
\newblock Large-reach memory management unit caches.
\newblock In {\em International Symposium on Microarchitecture}, pages
  383--394, 2013.

\bibitem{bock2019rip}
Carsten Bock, Ferdinand Brasser, David Gens, Christopher Liebchen, and
  Ahamd-Reza Sadeghi.
\newblock {RIP-RH}: Preventing rowhammer-based inter-process attacks.
\newblock In {\em Asia Conference on Computer and Communications Security},
  pages 561--572, 2019.

\bibitem{bonwick1994slab}
Jeff Bonwick.
\newblock The slab allocator: An object-caching kernel memory allocator.
\newblock In {\em USENIX summer}, 1994.

\bibitem{bosman2016dedup}
Erik Bosman, Kaveh Razavi, Herbert Bos, and Cristiano Giuffrida.
\newblock Dedup est machina: memory deduplication as an advanced exploitation
  vector.
\newblock In {\em IEEE Symposium on Security and Privacy}, pages 987--1004,
  2016.

\bibitem{brasser17can}
Ferdinand Brasser, Lucas Davi, David Gens, Christopher Liebchen, and Ahmad-Reza
  Sadeghi.
\newblock {CAn't} {Touch} {This}: Software-only mitigation against rowhammer
  attacks targeting kernel memory.
\newblock In {\em USENIX Security Symposium}, 2017.

\bibitem{cheng2018still}
Yueqiang Cheng, Zhi Zhang, Surya Nepal, and Zhi Wang.
\newblock {CATTmew}: Defeating software-only physical kernel isolation.
\newblock {\em IEEE Transactions on Dependable and Secure Computing}, 2019.

\bibitem{cojocar2020we}
Lucian Cojocar, Jeremie Kim, Minesh Patel, Lillian Tsai, Stefan Saroiu, Alec
  Wolman, and Onur Mutlu.
\newblock Are we susceptible to rowhammer? an end-to-end methodology for cloud
  providers.
\newblock In {\em IEEE Symposium on Security and Privacy}, May 2020.

\bibitem{cojocar2019exploiting}
Lucian Cojocar, Kaveh Razavi, Cristiano Giuffrida, and Herbert Bos.
\newblock Exploiting correcting codes: on the effectiveness of {ECC} memory
  against rowhammer attacks.
\newblock In {\em IEEE Symposium on Security and Privacy}, pages 55--71, 2019.

\bibitem{redblack}
Jonathan Corbet.
\newblock Trees ii: red-black trees.
\newblock \url{https://lwn.net/Articles/184495/}, 2006.

\bibitem{frigo_trrespass_2020}
Pietro Frigo, Emanuele Vannacci, Hasan Hassan, Victor van~der Veen, Onur Mutlu,
  Cristiano Giuffrida, Herbert Bos, and Kaveh Razavi.
\newblock {TRRespass}: Exploiting the many sides of target row refresh.
\newblock In {\em IEEE Symposium on Security and Privacy}, 2020.

\bibitem{gandhi2014badgertrap}
Jayneel Gandhi, Arkaprava Basu, Mark~D Hill, and Michael~M Swift.
\newblock Badgertrap: A tool to instrument x86-64 tlb misses.
\newblock {\em ACM SIGARCH Computer Architecture News}, 42(2):20--23, 2014.

\bibitem{gebai2018survey}
Mohamad Gebai and Michel~R Dagenais.
\newblock Survey and analysis of kernel and userspace tracers on linux: Design,
  implementation, and overhead.
\newblock {\em ACM Computing Surveys}, pages 1--33, 2018.

\bibitem{gruss2017another}
Daniel Gruss, Moritz Lipp, Michael Schwarz, Daniel Genkin, Jonas Juffinger,
  Sioli O'Connell, Wolfgang Schoechl, and Yuval Yarom.
\newblock Another flip in the wall of rowhammer defenses.
\newblock In {\em IEEE Symposium on Security and Privacy}, pages 245--261,
  2018.

\bibitem{gruss2016rowhammer}
Daniel Gruss, Cl{\'e}mentine Maurice, and Stefan Mangard.
\newblock Rowhammer.js: A remote software-induced fault attack in {JavaScript}.
\newblock In {\em Detection of Intrusions and Malware, and Vulnerability
  Assessment}, pages 300--321, 2016.

\bibitem{HP}
{HP, Inc.}
\newblock Hp moonshot component pack.
\newblock
  \url{https://support.hpe.com/hpsc/doc/public/display?docId=c04676483}, May
  2015.

\bibitem{intel}
{Intel, Inc.}
\newblock Intel 64 and {IA}-32 architectures software developer's manual
  combined volumes: 1, 2a, 2b, 2c, 3a, 3b and 3c.
\newblock October 2011.

\bibitem{intelecc}
{Intel, Inc.}
\newblock The role of ecc memory.
\newblock
  \url{https://www.intel.com/content/www/us/en/workstations/workstation-ecc-memory-brief.html},
  2015.

\bibitem{lpDDR4}
{JEDEC Solid State Technology Association.}
\newblock Low power double data rate 4 ({LPDDR4}).
\newblock \url{https://www.jedec.org/standards-documents/docs/jesd209-4b},
  2015.

\bibitem{kernelmap}
{Kernel.org}.
\newblock Virtual memory map with 4 level page tables (x86\_64).
\newblock \url{https://www.kernel.org/doc/Documentation/x86/x86_64/mm.txt},
  2009.

\bibitem{kim2020revisiting}
Jeremie~S Kim, Minesh Patel, A~Giray Yaglikci, Hasan Hassan, Roknoddin Azizi,
  Lois Orosa, and Onur Mutlu.
\newblock Revisiting rowhammer: An experimental analysis of modern dram devices
  and mitigation techniques.
\newblock In {\em International Symposium on Computer Architecture}, 2020.

\bibitem{kim2014flipping}
Yoongu Kim, Ross Daly, Jeremie Kim, Chris Fallin, Ji~Hye Lee, Donghyuk Lee,
  Chris Wilkerson, Konrad Lai, and Onur Mutlu.
\newblock Flipping bits in memory without accessing them: an experimental study
  of {DRAM} disturbance errors.
\newblock In {\em International Symposium on Computer Architecture}, page
  361–372, 2014.

\bibitem{konoth2018zebram}
Radhesh~Krishnan Konoth, Marco Oliverio, Andrei Tatar, Dennis Andriesse,
  Herbert Bos, Cristiano Giuffrida, and Kaveh Razavi.
\newblock Zeb{RAM}: comprehensive and compatible software protection against
  rowhammer attacks.
\newblock In {\em Operating Systems Design and Implementation}, pages 697--710,
  2018.

\bibitem{kurmus2014quantifiable}
Anil Kurmus, Sergej Dechand, and R{\"u}diger Kapitza.
\newblock Quantifiable run-time kernel attack surface reduction.
\newblock In {\em International Conference on Detection of Intrusions and
  Malware, and Vulnerability Assessment}, pages 212--234, 2014.

\bibitem{kurmus2011attack}
Anil Kurmus, Alessandro Sorniotti, and R{\"u}diger Kapitza.
\newblock Attack surface reduction for commodity os kernels: trimmed garden
  plants may attract less bugs.
\newblock In {\em Proceedings of the Fourth European Workshop on System
  Security}, pages 1--6, 2011.

\bibitem{kwong2020rambleed}
Andrew Kwong, Daniel Genkin, Daniel Gruss, and Yuval Yarom.
\newblock {RAMB}leed: Reading bits in memory without accessing them.
\newblock In {\em IEEE Symposium on Security and Privacy}, 2020.

\bibitem{lee2019twice}
Eojin Lee, Ingab Kang, Sukhan Lee, G~Edward Suh, and Jung~Ho Ahn.
\newblock {TWiCe}: preventing row-hammering by exploiting time window counters.
\newblock In {\em International Symposium on Computer Architecture}, pages
  385--396, 2019.

\bibitem{LENOVO}
{LENOVO, Inc.}
\newblock Row hammer privilege escalation lenovo security advisory.
\newblock \url{https://support.lenovo.com/au/en/product\_security/row\_hammer},
  August 2015.

\bibitem{DDR4}
{Micron, Inc.}
\newblock {DDR4 SDRAM Datasheet}.
\newblock \url{https://www.micron.com/products/dram/ddr4-sdram/}, 2015.

\bibitem{moscibroda2007memory}
Thomas Moscibroda and Onur Mutlu.
\newblock Memory performance attacks: Denial of memory service in multi-core
  systems.
\newblock In {\em USENIX Security Symposium}, 2007.

\bibitem{rowhammer}
Onur Mutlu.
\newblock Rowhammer.
\newblock
  \url{https://people.inf.ethz.ch/omutlu/pub/onur-Rowhammer-TopPicksinHardwareEmbeddedSecurity-November-8-2018.pdf},
  2018.

\bibitem{mutlu2019rowhammer}
Onur Mutlu and Jeremie~S Kim.
\newblock Rowhammer: A retrospective.
\newblock {\em IEEE Transactions on Computer-Aided Design of Integrated
  Circuits and Systems}, 2019.

\bibitem{park2020graphene}
Yeonhong Park, Woosuk Kwon, Eojin Lee, Tae~Jun Ham, Jung~Ho Ahn, and Jae~W Lee.
\newblock Graphene: Strong yet lightweight row hammer protection.
\newblock In {\em International Symposium on Microarchitecture}, pages 1--13,
  2020.

\bibitem{pessl2016drama}
Peter Pessl, Daniel Gruss, Cl{\'e}mentine Maurice, Michael Schwarz, and Stefan
  Mangard.
\newblock {DRAMA}: Exploiting {DRAM} addressing for cross-{CPU} attacks.
\newblock In {\em USENIX Security Symposium}, pages 565--581, 2016.

\bibitem{seaborndram}
Mark Seaborn.
\newblock How physical addresses map to rows and banks in dram.
\newblock
  \url{http://lackingrhoticity.blogspot.com.au/2015/05/how-physical-addresses-map-to-rows-and-banks.html},
  2015.

\bibitem{seaborn2015exploiting}
Mark Seaborn and Thomas Dullien.
\newblock Exploiting the {DRAM} rowhammer bug to gain kernel privileges.
\newblock In {\em Black Hat'15}, 2015.

\bibitem{seyedzadeh2016counter}
Seyed~Mohammad Seyedzadeh, Alex~K Jones, and Rami Melhem.
\newblock Counter-based tree structure for row hammering mitigation in {DRAM}.
\newblock {\em IEEE Computer Architecture Letters}, 16(1):18--21, 2016.

\bibitem{seyedzadeh2018mitigating}
Seyed~Mohammad Seyedzadeh, Alex~K Jones, and Rami Melhem.
\newblock Mitigating wordline crosstalk using adaptive trees of counters.
\newblock In {\em International Symposium on Computer Architecture}, pages
  612--623, 2018.

\bibitem{son2017making}
Mungyu Son, Hyunsun Park, Junwhan Ahn, and Sungjoo Yoo.
\newblock Making {DRAM} stronger against row hammering.
\newblock In {\em Design Automation Conference}, pages 1--6, 2017.

\bibitem{specint2017}
{Standard Performance Evaluation Corporation}.
\newblock Spec cpu 2017.
\newblock \url{https://www.spec.org}, 2017.

\bibitem{Nikto}
Chris Sullo.
\newblock \url{https://cirt.net/nikto}, 2012.

\bibitem{tatar2018throwhammer}
Andrei Tatar, Radhesh~Krishnan Konoth, Elias Athanasopoulos, Cristiano
  Giuffrida, Herbert Bos, and Kaveh Razavi.
\newblock Throwhammer: Rowhammer attacks over the network and defenses.
\newblock In {\em USENIX Annual Technical Conference}, 2018.

\bibitem{van2016drammer}
Victor van~der Veen, Yanick Fratantonio, Martina Lindorfer, Daniel Gruss,
  Cl{\'e}mentine Maurice, Giovanni Vigna, Herbert Bos, Kaveh Razavi, and
  Cristiano Giuffrida.
\newblock Drammer: Deterministic rowhammer attacks on mobile platforms.
\newblock In {\em ACM SIGSAC Conference on Computer and Communications
  Security}, pages 1675--1689, 2016.

\bibitem{van2018guardion}
Victor van~der Veen, Martina Lindorfer, Yanick Fratantonio,
  Harikrishnan~Padmanabha Pillai, Giovanni Vigna, Christopher Kruegel, Herbert
  Bos, and Kaveh Razavi.
\newblock Guardion: Practical mitigation of dma-based rowhammer attacks on arm.
\newblock In {\em International Conference on Detection of Intrusions and
  Malware, and Vulnerability Assessment}, pages 92--113. Springer, 2018.

\bibitem{wang2020dramdig}
Minghua Wang, Zhi Zhang, Yueqiang Cheng, and Surya Nepal.
\newblock Dramdig: A knowledge-assisted tool to uncover dram address mapping.
\newblock In {\em Design Automation Conference}, 2020.

\bibitem{wang2019safehidden}
Zhe Wang, Chenggang Wu, Yinqian Zhang, Bowen Tang, Pen-Chung Yew, Mengyao Xie,
  Yuanming Lai, Yan Kang, Yueqiang Cheng, and Zhiping Shi.
\newblock Safehidden: an efficient and secure information hiding technique
  using re-randomization.
\newblock In {\em USENIX Security Symposium}, pages 1239--1256, 2019.

\bibitem{wu2018CAT}
Xin-Chuan Wu, Timothy Sherwood, Frederic~T. Chong, and Yanjing Li.
\newblock Protecting page tables from rowhammer attacks using monotonic
  pointers in {DRAM} true-cells.
\newblock In {\em Architectural Support for Programming Languages and Operating
  Systems}, pages 645--657, 2019.

\bibitem{xenmap}
{xenbits.xen.org}.
\newblock source code (page.h).
\newblock
  \url{http://xenbits.xen.org/gitweb/?p=xen.git;a=blob;hb=refs/heads/stable-4.3;f=xen/include/asm-x86/x86_64/page.h},
  2009.

\bibitem{xiao2016one}
Yuan Xiao, Xiaokuan Zhang, Yinqian Zhang, and Radu Teodorescu.
\newblock One bit flips, one cloud flops: Cross-{VM} row hammer attacks and
  privilege escalation.
\newblock In {\em USENIX Security Symposium}, pages 19--35, 2016.

\bibitem{yauglikcci2021blockhammer}
Abdullah~Giray Ya{\u{g}}l{\i}k{\c{c}}{\i}, Minesh Patel, Jeremie~S Kim,
  Roknoddin Azizi, Ataberk Olgun, Lois Orosa, Hasan Hassan, Jisung Park,
  Konstantinos Kanellopoulos, Taha Shahroodi, Ghose Saugata, and Mutlu Onur.
\newblock Blockhammer: Preventing rowhammer at low cost by blacklisting
  rapidly-accessed dram rows.
\newblock In {\em High Performance Computer Architecture}, 2021.

\bibitem{zhangleveraging}
Zhenkai Zhang, Zihao Zhan, Daniel Balasubramanian, Bo~Li, Peter Volgyesi, and
  Xenofon Koutsoukos.
\newblock Leveraging {EM} side-channel information to detect rowhammer attacks.
\newblock In {\em IEEE Symposium on Security and Privacy}, 2020.

\bibitem{zhang2019telehammer}
Zhi Zhang, Yueqiang Cheng, Dongxi Liu, Surya Nepal, Zhi Wang, and Yuval Yarom.
\newblock Pthammer: Cross-user-kernel-boundary rowhammer through implicit
  accesses.
\newblock In {\em International Symposium on Microarchitecture}, 2020.

\end{thebibliography}

\end{document}